\documentclass[fleqn,usenatbib]{mnras}

\usepackage{graphicx}
\usepackage{amsmath}
\usepackage{amssymb}
\usepackage{url}
\usepackage{CJKutf8}
\usepackage{array}
\usepackage{threeparttable}
\usepackage[T1]{fontenc}
\usepackage{tabularx}
\usepackage{tipa}
\usepackage{scalerel,tikz}

\usetikzlibrary{svg.path}
\definecolor{orcidlogocol}{HTML}{A6CE39}
\tikzset{orcidlogo/.pic={
 \fill[orcidlogocol] svg{M256,128c0,70.7-57.3,128-128,128C57.3,256,0,198.7,0,128C0,57.3,57.3,0,128,0C198.7,0,256,57.3,256,128z};
 \fill[white] svg{M86.3,186.2H70.9V79.1h15.4v48.4V186.2z}
 svg{M108.9,79.1h41.6c39.6,0,57,28.3,57,53.6c0,27.5-21.5,53.6-56.8,53.6h-41.8V79.1z M124.3,172.4h24.5c34.9,0,42.9-26.5,42.9-39.7c0-21.5-13.7-39.7-43.7-39.7h-23.7V172.4z}
 svg{M88.7,56.8c0,5.5-4.5,10.1-10.1,10.1c-5.6,0-10.1-4.6-10.1-10.1c0-5.6,4.5-10.1,10.1-10.1C84.2,46.7,88.7,51.3,88.7,56.8z};
}}
\newcommand\orcidicon[1]{\href{https://orcid.org/#1}{\mbox{\scalerel*{
\begin{tikzpicture}[yscale=-1,transform shape]
\pic{orcidlogo};
\end{tikzpicture}
}{|}}}}

\makeatletter
\newcommand*{\@rowstyle}{}
\newcommand*{\rowstyle}[1]{
\gdef\@rowstyle{#1}%
\@rowstyle\ignorespaces%
}
\newcolumntype{=}{
>{\gdef\@rowstyle{}}%
}
\newcolumntype{+}{
>{\@rowstyle}%
}
\newcolumntype{C}[1]{>{\centering\arraybackslash}p{#1}}
\makeatother

\newcommand{\help}[1]{{#1}}
\newcommand{\resp}[1]{{#1}}

\newcommand{\lya}{\ifmmode\mathrm{Ly}\alpha\else{}Ly$\alpha$\fi\:}
\newcommand{\lyb}{\ifmmode\mathrm{Ly}\beta\else{}Ly$\beta$\fi}
\newcommand{\igm}{\ifmmode\mathrm{IGM}\else{}IGM\fi}
\newcommand{\lae}{\ifmmode\mathrm{LAE}\else{}LAE\fi}
\newcommand{\h}{\ifmmode\mathrm{H}\else{}H\fi}
\newcommand{\hi}{\ifmmode\mathrm{H\,{\scriptscriptstyle I}}\else{}H\,{\scriptsize I}\:\fi}
\newcommand{\heii}{\ifmmode\mathrm{He\,{\scriptscriptstyle II}}\else{}He\,{\scriptsize II}\fi}
\newcommand{\heiii}{\ifmmode\mathrm{He\,{\scriptscriptstyle III}}\else{}He\,{\scriptsize III}\fi}
\newcommand{\ciii}{\ifmmode\mathrm{C\,{\scriptscriptstyle III]}}\else{}C\,{\scriptsize III]}\fi}
\newcommand{\oiii}{\ifmmode\mathrm{O\,{\scriptscriptstyle III}}\else{}O\,{\scriptsize III}\fi}
\newcommand{\aliii}{\ifmmode\mathrm{Al\,{\scriptscriptstyle III}}\else{}Al\,{\scriptsize III}\fi}
\newcommand{\mgii}{\ifmmode\mathrm{Mg\,{\scriptscriptstyle II}}\else{}Mg\,{\scriptsize II}\fi}
\newcommand{\fe}{\ifmmode\mathrm{Fe}\else{}Fe\fi}
\newcommand{\nv}{\ifmmode\mathrm{N\,{\scriptscriptstyle V}}\else{}N\,{\scriptsize V}\fi}
\newcommand{\niv}{\ifmmode\mathrm{N\,{\scriptscriptstyle IV]}}\else{}N\,{\scriptsize IV]}\fi}
\newcommand{\cii}{\ifmmode\mathrm{C\,{\scriptscriptstyle II}}\else{}C\,{\scriptsize II}\fi}
\newcommand{\civ}{\ifmmode\mathrm{C\,{\scriptscriptstyle IV}}\else{}C\,{\scriptsize IV}\fi}
\newcommand{\siv}{\ifmmode\mathrm{Si\,{\scriptscriptstyle IV}}\else{}Si\,{\scriptsize IV}\fi}
\newcommand{\siii}{\ifmmode\mathrm{Si\,{\scriptscriptstyle II}}\else{}Si\,{\scriptsize II}\fi}
\newcommand{\siiii}{\ifmmode\mathrm{Si\,{\scriptscriptstyle III]}}\else{}Si\,{\scriptsize III]}\fi}
\newcommand{\ovi}{\ifmmode\mathrm{O\,{\scriptscriptstyle VI}}\else{}O\,{\scriptsize VI}\fi}
\newcommand{\sioiv}{\ifmmode\mathrm{Si\,{\scriptscriptstyle IV}\,\plus O\,{\scriptscriptstyle IV]}}\else{}Si\,{\scriptsize IV}\,+O\,{\scriptsize IV]}\fi}
\newcommand{\Msun}{\rm M_\odot}

\newcommand{\appropto}{\mathrel{\vcentre{
			\offinterlineskip\halign{\hfil$##$\cr
				\propto\cr\noalign{\kern2pt}\sim\cr\noalign{\kern-2pt}}}}}

\newcommand{\lsim}{\mathrel{\rlap{\lower4pt\hbox{\hskip1pt$\sim$}}
        \raise1pt\hbox{$<$}}}
\newcommand{\gsim}{\mathrel{\rlap{\lower4pt\hbox{\hskip1pt$\sim$}}
        \raise1pt\hbox{$>$}}}
\newcommand{\Rom}[1]{\uppercase\expandafter{\romannumeral #1}}
\newcommand{\rom}[1]{\lowercase\expandafter{\romannumeral #1}}

\newcommand{\oneh}{h^{-1}}

\newcommand{\music}{{$\textsc{MUSIC}$}\:}
\newcommand{\dorcha}{{$\textsc{Dorcha}$}}
\newcommand{\D}[1]{\textsc{Dorcha}\_\textsc{#1}}

\pdfoutput=1

\title[\textsc{Dorcha} - DMO Milky Way Analogue zooms]{The \textsc{Dorcha} suite: nature, nurture, and the phase space distribution of the Milky Way's high redshift progenitors today} 
\author[Balu et al.]{Sreedhar Balu\orcidicon{0000-0002-5281-5151}$^{1, 2, 5}$\thanks{E-mail:bsreedhar@us.es},
Chris Power\orcidicon{0000-0002-4003-0904}$^{3, 5}$,
Kris Walker\orcidicon{0000-0003-2128-1289}$^{3, 5}$, \&
J. Stuart B. Wyithe\orcidicon{0000-0001-7956-9758}$^{4, 5}$
\\
$^{1}$Facultad de Físicas, Multidisciplinary Unit for Energy Science, Universidad de Sevilla, 41012, Seville, Spain\\
$^{2}$School of Physics, University of Melbourne, Parkville, VIC 3010, Australia\\
$^{3}$International Centre for Radio Astronomy Research, M468, University of Western Australia, 35 Stirling Hwy, Perth, WA 6009, Australia\\
$^{4}$Research School of Astronomy and Astrophysics, Australian National University, Canberra, ACT 2611, Australia\\
$^{5}$ARC Centre of Excellence for All Sky Astrophysics in 3 Dimensions (ASTRO 3D)
}

\date{Accepted XXX. Received YYY; in original form ZZZ}

\pubyear{\the\year{}}

\begin{document}

\label{firstpage}
\pagerange{\pageref{firstpage}--\pageref{lastpage}}
\maketitle
\begin{abstract}
\noindent
\help{Where in the present-day Milky Way should we search for the remnants of its earliest stars? We address this question using the \dorcha~(Gaelic for \textit{Dark}; \textsc{DUR-uh-khuh}) suite: a set of 25 high-resolution, dark-matter-only cosmological zoom-in simulations of Milky Way analogue (MWA) haloes evolved to $z=0$. Of these, 15 are isolated and the rest are in pairs, similar to the MW and M31. By identifying and tagging the most bound material in high-redshift ($z\geq5$) progenitor haloes -- those likely to host early star formation -- we track the present-day phase-space distribution of this ancient component. We find that this material is highly centrally concentrated at $z=0$, with 90 -- 100 per cent residing within $r \lesssim 15\,h^{-1}\,\mathrm{kpc}$. It exhibits steep density profiles ($\rho\propto\,r^{-4}$), low velocity dispersions ($\sigma_r / \sigma_{\rm max} \lesssim 0.6$), and radially biased orbits ($\beta \gtrsim 0.5$ for $r \gtrsim 0.1\,R_{200}$), consistent with a relaxed, centrally embedded population. These results hold across haloes with diverse formation histories and environments, suggesting that the dynamical signature of early progenitors is robust to later mergers and interactions.
Our findings imply that the fossil record of the first generations of stars -- including Population III and extremely metal-poor stars -- should be sought in the innermost regions of the Milky Way, where they retain distinctive kinematic imprints. {While these stellar populations may overlap, we caution that low metallicity does not uniquely identify ancient stars, nor vice versa.} The \dorcha~suite thus provides a physically motivated baseline for interpreting observations from Galactic Archaeology surveys targeting the bulge and inner halo.}
\end{abstract}

\begin{keywords}
Galaxy: general -- kinematics and dynamics -- evolution -- Cosmology: dark matter
\end{keywords}

\section{Introduction} 
\label{sec:intro}

 

The last several years have witnessed a remarkable  advance in astronomers' capacity to measure the properties of statistical samples of galaxies at high redshifts ($z\gtrsim5$) and to characterise their larger-scale environment, with important consequences for our understanding of galaxy formation and evolution in the early Universe. This is most strikingly apparent in the results coming from \textsc{JWST}, which have revealed unexpectedly high inferred stellar masses of high redshift massive galaxies \citep[e.g.][]{boylan-kolchinStressTestingLCDM2023, caseyCOSMOSWebIntrinsicallyLuminous2024}, in apparent tension with theoretical models for early galaxy formation \citep[cf.][]{starkObservationsFirstGalaxies2026}. Forthcoming direct measurements of the cosmic distribution of neutral atomic hydrogen via the redshifted 21-cm line \citep[e.g.][]{BibleReview, StuartReview} by the Square Kilometre Array radio telescope \citep[see, e.g,][]{Mellema2013}
promise a similar transformation in our understanding of cold gas the high redshift Universe. These measurements will reveal the conditions under which early generations of galaxies formed and evolved, and the sources of the ionizing radiation background that led to cosmological reionization \citep[e.g.][]{BarkanaReview}.  

However, we can access a complementary understanding of the conditions of galaxy formation in the early Universe by studying its artefacts in the present-day ($z=0$) Universe. Galactic Archaeology \citep[e.g.][]{freemanNewGalaxySignatures2002, Frebel_Norris_2015} targets the Local Group -- the Milky Way and Andromeda (M31), their retinue of satellites and globular clusters, as well as stellar streams and haloes -- and dissects the intrinsic properties (e.g. stellar ages and metallicities), spatial structure, kinematics, and inferred dynamics of its constituent systems to characterise their formation and assembly histories. This motivates ongoing Galactic mapping and archaeology surveys such as GALAH \citep[cf.][]{desilvaGALAHSurveyScientific2015}, Gaia \citep[cf.][]{Gaia_2016} and APOGEE \citep[cf.][]{majewskiApachePointObservatory2017}, and forthcoming surveys and facilities such as 4MOST \citep[cf.][]{dejong4MOSTProjectOverview2019} and LSST \citep[cf.][]{LSST_2019}. 

These surveys contain a wealth of data -- positions, kinematics, ages, and metallicities -- from which detailed maps of the oldest stars around the Milky Way and in the Local Group have been constructed \citep[cf.][]{bland-hawthornGalaxyContextStructural2016a}. Age-dating via elemental abundances and stellar astro-seismic signatures reveal that the oldest stars formed within a few hundred million years after the Big Bang \citep[e.g.][]{freemanNewGalaxySignatures2002, kochKinematicChemicalConstraints2008, desilvaGALAHSurveyScientific2015, tolstoyStarFormationHistoriesAbundances2009}. These old stars are found within multiple components -- the stellar halo \citep{carolloStructureKinematicsStellar2010, helmiStreamsSubstructuresEarly2020}; the Galactic bulge \citep{barbuyChemodynamicalHistoryGalactic2018}; dwarf galaxies \citep{tolstoyStarFormationHistoriesAbundances2009, mcconnachieMaunakeaSpectroscopicExplorer2014}; and globular clusters \citep{pritzlComparisonElementalAbundance2005} -- and they are observed to trace distinct spatial and kinematic distributions. For example, the old stellar halo population is concentrated in an inner component; overlaps with the Galactic bulge; and it follows a preferentially radial orbital distribution, in contrast to the outer component \citep[cf.][]{bland-hawthornGalaxyContextStructural2016a}. \help{Although metallicity is often used as a proxy for age, especially in Galactic Archaeology, the two are not strictly equivalent. Extremely metal-poor stars can form after the first generations of star formation, and not all old stars necessarily have low metallicities \citep[see e.g.][]{whiteWhereAreFirst2000,bergemannGaiaESOSurveyRadial2014, el-badryWhereAreMost2018}. This distinction is important for interpreting observational data.}

\help{Because of the hierarchical and inside-out growth of dark matter (DM) haloes, we expect the inner regions of the Milky Way to host most of its oldest components \citep[e.g.]{diemandDistributionKinematicsEarly2005, cooperGalacticStellarHaloes2010, deasonEatingHabitsMilky2016}. Observational campaigns targeting old (i.e. formed at $z\geq5$) and metal-poor ([Fe/H]$<-2.5$) stars -- which serve as luminous tracers of these components -- have overlooked the inner regions of the Milky Way because of the challenges in disentangling the effects of severe dust obscuration and the dominance of the more numerous metal-rich components. 
We expect that the number of metal-poor and old stars in the innermost parts of the Milky Way to be negligible fraction of the total number of stars present; \resp{ for example, \citet{starkenburgOldestMostMetalpoor2017} used the APOSTLE simulation to show that $\lesssim 2\%$ of the total stellar population in the inner Galaxy is very metal-poor ([Fe/H]$<-2.5$) and old (formed at $z>6.9$). }For this reason, we do not expect a random survey to separate the old components \citep{nessARGOSIVKinematics2013} from the metal-rich stars in this environment, whereas targeted surveys such as EMBLA \citep[cf. ][]{howesGaiaESOSurveyMost2014, howesEMBLASurveyMetalpoor2016} and COMBS \citep[cf.][]{luceyCOMBSSurveyChemical2019, luceyCOMBSSurveyII2021, luceyCOMBSSurveyIII2021} have detected extremely metal-poor stars within the inner parts of the Milky Way. Interestingly, the Pristine Inner Galaxy Survey (PIGS) has reported low-metallicity stars in the Galaxy's inner bulge having distinct kinematic and dynamical features when separating their samples based on the metallicity \citep[cf.][]{arentsenPristineInnerGalaxy2020, ardern-arentsenPristineInnerGalaxy2024}
}

Given these observed (present-day) properties of the Milky Way's oldest stars, we can ask what inferences we can about the assembly history of the Milky Way and the conditions under which its high redshift progenitors formed and evolved. In particular, we can ask how sensitive these inferences are to (1) the peculiarities of the Milky Way -- such as the relatively isolated nature of the Local Group, and the proximity of Andromeda -- and to (2) the assumptions we make about how galaxy formation proceeds at early times. 
\medskip

Here we use cosmological $N$-body simulations to address the first of these questions, and address the second in forthcoming work.
\resp{Theoretical simulations provide a powerful interpretive framework to connect the present-day observable properties of the Milky Way's stellar populations to its assembly history, and various approaches have been used. These span non-cosmological $N$-body simulations of the tidal evolution of satellite galaxies \citep[e.g.][]{bullockTracingGalaxyFormation2005}; cosmological $N$-body simulations \citep[e.g.][]{whiteWhereAreFirst2000,madauDarkMatterSubhalos2008,springelAquariusProjectSubhalos2008,gaoEarliestStarsTheir2010} coupled to semi-analytical galaxy formation models \citep[e.g.][]{cooperGalacticStellarHaloes2010,deasonEatingHabitsMilky2016}; and cosmological hydrodynamical galaxy formation models \citep[e.g.][]{starkenburgOldestMostMetalpoor2017,monachesiAurigaStellarHaloes2019,santistevanFormationTimesBuilding2020}.}

\resp{These complementary approaches provide insights into how the Milky Way, its satellites, and stellar streams and halo, have evolved from high redshift to the present-day, and inform our current theoretical picture. Early work using non-cosmological models of disrupting satellite galaxies in a Galactic potential showed that the stellar halo of the Milky Way most likely grew from the inside-out, with an old, centrally concentrated, stellar component of stars on preferentially radial orbits  \citep{bullockTracingGalaxyFormation2005}. Cosmological $N$-body simulations \citep[e.g.][]{whiteWhereAreFirst2000,Madau_2008,springelAquariusProjectSubhalos2008, gaoEarliestStarsTheir2010} showed that such profiles follow naturally if we assume that these old stars are associated with high-$\sigma$ peaks\footnote{We can characterise overdensities in terms of $\nu\sigma(M,z)$, where $\nu$ quantifies the rarity of the overdensity, $\sigma(M,z)$ is the root-mean-square fluctuation of the linearly extrapolated matter density field smoothed with a top-hat filter on a mass scale $M$, and $z$ is the redshift. A 1-$\sigma$ peak corresponds to the characteristic mass $M_\ast(z)$ such that $\sigma(M_\ast,z)=1.69$.} in the early Universe. However, identifying these stars today is not straightforward - cosmological galaxy formation simulations reveal that ancient or extremely metal‑poor stars contribute a negligible fraction within the inner Milky Way \citep[e.g.][]{starkenburgOldestMostMetalpoor2017, santistevanFormationTimesBuilding2020} - and requires us to leverage the distinct physical properties of the population.}

The spatial and kinematic distributions of this material in the present day Milky Way should show a number of distinctive features -- compared to the overall mass distribution, it will be more centrally concentrated; fall off more steeply with radius; have a lower velocity dispersion; and move on more radial orbits \citep[e.g.][]{diemandDistributionKinematicsEarly2005, hortaProtogalaxyMilkyWaymass2023,reyEDGEEmergenceDwarf2025}. The origin of this material can be traced to a few massive dwarfs (peak $M_\ast\geq10^8 - 10^9 M_\odot$), which dominate the inner halo as a result of early dynamical‑friction and tidal disruption, and numerous low‑mass accretions, which populate the outer halo, typically as coherent streams \citep[cf.][]{hortaProtogalaxyMilkyWaymass2023,geninaEdgeRelationStellar2023}.
Metallicity contains important information about the number and masses of progenitors - steep negative [Fe/H] gradients indicate halos built by few, massive mergers, whereas flatter gradients reflect many lower‑mass contributors \citep[e.g.][]{monachesiAurigaStellarHaloes2019} - but there are crucial caveats. Inner‑halo chemistry is degenerate because in‑situ, merger‑driven star formation can overlap with accreted material in [Fe/H]–[$\alpha$/Fe], meaning that ages and multi‑element diagnostics are needed to disentangle origins \citep[e.g.][]{reyEDGEEmergenceDwarf2025}.

The consistency of the present radial mass density profiles of material associated with high redshift high-$\sigma$ peaks with profiles of the Milky Way's stellar bulge and halo, dwarf galaxies, and globulars clusters \citep[cf.][]{diemandDistributionKinematicsEarly2005} is highly suggestive that we can map present-day phase space structure to its high-redshift progenitors. Indeed, if old stellar material has been relatively recently accreted by the Milky Way, it may continue to be a coherent substructure in phase space \citep[e.g.][]{knebeMappingSubstructuresDark2005,geninaEdgeRelationStellar2023, reyMEGATRONHowFirst2025a}, but this becomes less likely the longer it orbits in the Milky Way's time-varying gravitational potential \citep[e.g.][]{amaranteGastroLibrarySimulated2022, moriMetallicityDistributionsHalo2024,thomasHowWellCan2025a}.

\medskip

In this paper, we use our new \dorcha\footnote{From Irish Gaelic, \textit{dorcha} translates to English as \textit{dark}; it is pronounced \textipa{["d\textopeno\:r\textschwa x\textschwa]}; roughly \textit{DUR-uh-khuh}.}~suite of cosmological $N$-body simulations to study what we can learn about the Milky Way's progenitors in the early Universe and its subsequent assembly history from old stellar populations at $z=0$. We do this using a statistical sample of 25 Milky Way Analogues (MWAs), selected to have masses and local environments comparable to the Milky Way; this sample consists of 15 isolated systems and 5 pairs. We can track MWA progenitors as early as $z=25$ and tag the most bound material in the structures capable of supporting atomic hydrogen cooling between $5\leq\,z\leq\,25$ as potential sites for high redshift star formation; this material provides us with dynamical tracers of the $z=0$ old stellar populations, which we note may encompass both metal-free and metal-poor stars. Our suite allows us to assess the degree to which assembly history and the close proximity of a massive neighbour is likely to influence the $z=0$ phase space distribution of the oldest stars in the Milky Way. For example, we expect the Milky Way to have a quiescent merging history, given the relatively early time when its disc was likely assembled \citep[e.g.][]{bland-hawthornGalaxyContextStructural2016a, belokurovDawnTillDisc2022}; how much of an impact does the plausible range of a MWA's assembly histories have on the old stellar population's phase space? Similarly, how is this phase space influenced if it co-evolves with a pair of galaxies like the Milky Way and Andromeda? 

We note that the \textsc{ELVIS} suite \citep[cf.][48 MWAs with half in pairs]{garrison-kimmelELVISExploringLocal2014}; the  \textsc{Caterpillar} suite \citep[cf.][30 MWAs]{griffenTracingFirstStars2018a}; and the \textsc{Milky Way-est} suite \citep[cf.][20 MWAs]{buch_milky_2024} have adopted a similar approach to us in using cosmological $N$-body simulations. An important point of difference is that the \dorcha~suite is a core part of our upcoming \textsc{Solas} suite\footnote{The Gaelic Irish word \textit{solas} means \textit{light}.} of simulations which will create hydrodynamical galaxy formation counterparts of our $N$-body haloes. By comparing and contrasting these simulations, we will address the second of the questions posed above (i.e. \emph{How sensitive are our inferences to the assumptions we make about how galaxy formation proceeds at early times?}). The \dorcha~suite helps us in developing a heuristic viewpoint that is not tied to specific models of star formation, feedback, and metal enrichment processes.

\medskip

In the following sections, we outline our methodological approach (\S~\ref{sec:methodology}), including a description of our simulations and analysis tools; we present our results (\S~\ref{sec:results}), including a visual impression of the \dorcha~suite, spherically averaged profiles, phase space structure, and orbital distributions; and we provide a summary of our conclusions (\S~\ref{sec:conclusions}). Unless otherwise specified, we use comoving units.

\section{Methodology}
\label{sec:methodology}
In this section, we present details on how we have run our simulations suite and our approach to halo finding and merger tree construction (\S~\ref{ssec:sims}); we describe how we selected the MWA candidates for resimulation (\S~\ref{ssec:halo_selection}); and we outline our approach to identifying material associated with high redshift haloes that might be plausible sites of early star formation (\S~\ref{ssec:highz_sf}). This is the material whose spatial and kinematic properties we quantify in the next section.

\vspace{-4mm}

\subsection{Simulations}
\label{ssec:sims}
\begin{table*}
\caption{\label{tab:sim_info}Details of all 25 \dorcha~haloes. \D{01-15} are isolated systems and \D{15-20} comprise of paired systems. $M_{200}$, $R_{200}$, and $V_{200}$ correspond to the virial masses, radii, and circular velocities of the haloes at $z$=0; $N_{\rm subs}$ is the number of subhaloes associated with the halo at $z$=0; $z_{\rm init}$ is the redshift at which the main progenitor of the halo was first identified in the \textsc{SubFind-HBT} catalogues; $z_{10}$ which is the redshift at which 10 per cent of the halo's $z$=0 mass was in place ; and $z_{\rm form}\equiv z_{50}$ is the formation redshift, which by convention is when 50 per cent of the main branch mass was in place. \resp{Our subhalo mass resolution, with a minimum of 32 particles, is $5.04\times 10^6~\oneh\Msun$.} We also highlight which haloes can be considered early or late forming (kind).}
\centering
\begin{tabularx}{\textwidth} { 
   >{\raggedright\arraybackslash}X 
   >{\centering\arraybackslash}X 
   >{\centering\arraybackslash}X 
   >{\centering\arraybackslash}X 
   >{\centering\arraybackslash}X 
   >{\centering\arraybackslash}X 
   >{\centering\arraybackslash}X 
   >{\centering\arraybackslash}X 
   >{\centering\arraybackslash}X  }
 \hline
 \vspace{0.1cm}\\
Halo & $M_{200}$ & $R_{200}$ & $V_{200}$  & $N_{subs}$ & $z_{\rm init}$  & $z_{10}$ & $z_{50}$ & kind \\
    & {\fontsize{6}{8}\selectfont $(\times 10^{12}~\oneh\:\Msun)$} & {\fontsize{6}{8}\selectfont $(\oneh\:kpc)$}  & {\fontsize{6}{8}\selectfont (km s$^{-1}$)} & $(z=0)$ & & & $(\equiv  z_{\rm form})$  &  \\ 
\vspace{0.1cm}\\
\D{01} & 1.47 & 185.02 & 185.0 & 2867 & 28.96 & 3.23  & 0.28 & late\\
\vspace{0.5mm}\\
\D{02} & 1.46 & 184.46 & 184.45 & 2660 & 27.97 & 4.75  & 2.22 & early\\
\vspace{0.5mm}\\
\D{03} & 1.22 & 173.68 & 173.67 & 2559 & 28.96 & 3.7  & 1.52 & --\\
\vspace{0.5mm}\\
\D{04} & 1.31 & 177.99 & 177.97 & 2194 & 22.82 & 2.71  & 1.06 & --\\
\vspace{0.5mm}\\
\D{05} & 1.08 & 166.75 & 166.72 & 2092 & 23.58 & 3.88  & 1.69 & early\\
\vspace{0.5mm}\\
\D{06} & 1.16 & 171.04 & 171.03 & 1872 & 26.17 & 4.21  & 1.93 & early\\
\vspace{0.5mm}\\
\D{07} & 1.06 & 165.7 & 165.68 & 2261 & 22.06 & 3.79  & 1.56 & --\\
\vspace{0.5mm}\\
\D{08} & 0.83 & 153.0 & 152.98 & 1737 & 13.56 & 2.33  & 1.17 & --\\
\vspace{0.5mm}\\
\D{09} & 0.72 & 146.04 & 146.02 & 1469 & 25.3 & 4.12  & 1.38 & --\\
\vspace{0.5mm}\\
\D{10} & 1.46 & 184.49 & 184.47 & 3135 & 22.82 & 3.93  & 1.72 & early\\
\vspace{0.5mm}\\
\D{11} & 1.03 & 164.45 & 164.44 & 1661 & 22.06 & 3.85  & 1.41 & --\\
\vspace{0.5mm}\\
\D{12} & 0.95 & 160.08 & 160.07 & 1837 & 27.07 & 3.97  & 1.36 & --\\
\vspace{0.5mm}\\
\D{13} & 0.73 & 146.7 & 146.68 & 1342 & 28.96 & 3.69  & 1.37 & --\\
\vspace{0.5mm}\\
\D{14} & 0.86 & 154.5 & 154.48 & 1353 & 22.06 & 4.26  & 1.28 & --\\
\vspace{0.5mm}\\
\D{15} & 0.65 & 140.62 & 140.6 & 1759 & 22.82 & 3.16  & 1.32 & --\\
\vspace{0.5mm}\\
\D{16A} & 1.57 & 189.1 & 189.07 & 3207 & 26.14 & 3.35  & 1.33 & --\\
\vspace{0.5mm}\\
\D{16B} & 1.57 & 188.89 & 188.87 & 3001 & 22.82 & 3.73  & 1.23 & --\\
\vspace{0.5mm}\\
\D{17A} & 1.08 & 166.72 & 166.7 & 2301 & 25.3 & 2.66  & 0.89 & late\\
\vspace{0.5mm}\\
\D{17B} & 0.75 & 147.42 & 147.41 & 2630 & 23.58 & 2.86  & 1.13 & --\\
\vspace{0.5mm}\\
\D{18A} & 1.07 & 166.32 & 166.3 & 2521 & 27.07 & 3.4  & 0.63 & late\\
\vspace{0.5mm}\\
\D{18B} & 0.89 & 156.7 & 156.68 & 1702 & 24.43 & 3.48  & 1.48 & --\\
\vspace{0.5mm}\\
\D{19A} & 1.01 & 163.2 & 163.18 & 1867 & 28.96 & 2.86  & 2.21 & early\\
\vspace{0.5mm}\\
\D{19B} & 0.93 & 158.71 & 158.7 & 2051 & 23.61 & 3.3  & 0.66 & late\\
\vspace{0.5mm}\\
\D{20A} & 0.95 & 160.07 & 160.06 & 2358 & 28.96 & 3.09  & 1.0 & late\\
\vspace{0.5mm}\\
\D{20B} & 0.91 & 157.31 & 157.29 & 2087 & 22.82 & 3.95  & 1.33 & --\\
\vspace{0.5mm}\\
\hline
\end{tabularx}
\end{table*}

\subsubsection{Cosmological Parameters}
\resp{We use the best-fit \cite{planckcollaborationPlanck2015Results2016} cosmological parameters, specifically the \texttt{TT,TE,EE+lowP+lensing} column of their Table 4.} The key parameters are, respectively, the matter, dark energy, and baryon density parameters; the present-day Hubble parameter, the mass variance normalisation; and the spectral index, with values of $(\Omega_{\rm m}, \Omega_{\Lambda}, \Omega_{\rm b}, H_{0}, \sigma_{8}, n_{\rm s}) = (0.3121, 0.6879, 0.0491,  67.51, 0.815, 0.9653)$. 

\vspace{-4mm}

\subsubsection{Simulation Code}
We use \textsc{arepo}
\citep{Arepo2010, weinbergerAREPOPublicCode2020} to run our cosmological $N$-body parent and zoom simulations. \textsc{arepo} employs a hybrid Tree-PM approach to estimate the gravitational force, splitting the calculation into a long-range component, which is solved via a particle mesh (PM) approach, and a short-range component, which is solved via a hierarchical oct-tree approach \citep[cf.][]{BarnesHut1986,GADGET}. Each simulation is run from a starting redshift of $z=99$ to $z=0$, and we use \music~\citep{hahnMultiscaleInitialConditions2011} to generate our initial conditions (ICs). 

\vspace{-4mm}

\subsubsection{The Simulation Suite}
Our parent simulation, \D{00}, is a uniform resolution periodic volume of side length $L$=100 $\oneh$Mpc containing $N$=$512^3$ particles (i.e. \music~input parameter \texttt{levelmin=9}), each of mass $m_\mathrm{p} \simeq 6.45\times10^{8}~\oneh\Msun$. We use a fixed comoving softening length of $\epsilon_{\rm f}= 0.04\times\ell$ \citep{Ludlow2019}, where $\ell=L/N$ is the mean inter-particle separation. We recast $\ell$ in the form, 
\begin{equation}
\ell=\left(m_\text{part}/\Omega_\text{m}\rho_\text{crit}\right)^{1/3},
\end{equation}
where $m_\text{part}$ is the particle mass, $\Omega_\text{m}$ is the matter density parameter, and $\rho_\text{crit}$ is the critical density; this is convenient when evaluating softenings for zoom simulations. This gives $\epsilon_{\rm f}=10~\oneh\rm{kpc}$. 

From the \D{00} halo catalogue at $z$=0, we build a list of candidate haloes for resimulation based on a set of criteria, which we set out in \S~\ref{ssec:halo_selection}, to select the hosts of Milky Way Analogues (MWAs). Our candidates consist of 15 isolated systems (\D{01-15}) and 10 in pairs (\D{16-20}). We select all the particles within some threshold radius $r_{\rm cut}$ at $z=0$, and use their positions at $z=99$ (in the ICs) to define the spatial extent of the Lagrangian volume to be resimulated. For our isolated MWAs, $r_{\rm cut}=4\times R_{\rm 200}$ where $R_{\rm 200}$ is the virial radius, while $r_{\rm cut}=1.5~\oneh\rm{Mpc}$ relative to their centre of mass for our paired MWAs. In all cases, we add a buffer of $0.5~\oneh\rm{Mpc}$ along each Cartesian dimension at $z=99$. The ICs of the MWA zoom simulations have an increased mass resolution of 4096 times relative to \D{00} (i.e. \music~input parameter \texttt{levelmax=13}). This corresponds to a particle mass in the zoom region of $1.57\times 10^5~\oneh\Msun$ for the \dorcha~suite of zoom simulations. The associated gravitational force softening of particles in the zoom region is $\epsilon_{\rm f}=50~\oneh\rm{pc}$. 

\vspace{-4mm}

\subsubsection{Halo Finding \& Merger Trees}
We use the version of \textsc{SubFind-HBT} in \textsc{GADGET-4} \citep{Gadget4} code to construct our halo catalogues. Haloes correspond to Friends-of-Friends (hereafter FoF) groups assuming a linking length of $b=0.2\times\ell$. We set a lower limit of \resp{32 particles for FoF groups as well as the subhaloes}, which gives us a minimum halo mass of $\mathrm{M}_\text{h,min}^\text{parent}\simeq2.06\times10^{10}~\oneh\Msun$ in the \D{00} parent simulation, and $5.04\times 10^6~\oneh\Msun$ in the \D{01-20} zoom simulations.
 
For the purposes of this paper, we treat FoF groups (\textsc{Groups} in \textsc{SubFind-HBT}) as haloes, which includes both gravitationally bound and unbound material associated with the halo. Our analysis can be applied equally to the first-ranked (central) subhalo within the FoF group (from \textsc{SubHaloes} in \textsc{SubFind-HBT}), which includes only gravitationally bound material associated with the smooth background halo. However, our inclusive definition means that we maximise the volume within which we might expect to find the remnants of progenitor haloes that hosted high redshift star formation. 

Note that in our analysis we use the values of the virial mass, $M_{200}$, and virial radius, $R_{200}$ computed by \textsc{SubFind-HBT}; these values are defined such that the mean enclosed density within $R_{200}$ at redshift $z$ is 200 times the critical density, $\rho_{\rm crit}(z)=3H^2(z)/8\pi G$, where $H(z)$ is the Hubble parameter and $G$ is the gravitational constant.

\textsc{SubFind-HBT} allows us to reliably track haloes and subhaloes across snapshots. This takes advantage of the decomposition of FoF haloes into their constituent central subhalo plus associated subhaloes -- material associated with a given subhalo is tracked forward in time across snapshots explicitly as part of the halo finding process, from the point at which it was first identified as a distinct dynamically self-consistent structure. This reduces the ambiguity when linking haloes and subhaloes across snapshots and consequently produces more reliable merger trees (cf. \citealt{chandro-gomezAccuracyDarkMatter2025}; also see Appendix B of \citealt{forouharmorenoAssessingSubhaloFinders2025}). 

\vspace{-4mm}

\subsubsection{Snapshot Cadence}
We use the same snapshot cadence for both parent and zoom simulations -- 123 snapshots between $z=30$ and $z=0$.
\begin{itemize}
\item There are 50 snapshots evenly spaced in $\log(a)$, the natural logarithm of the scalefactor, between $z=30$ and $5$; this corresponds to a snapshot cadence with a minimum(maximum) temporal separation between consecutive snapshots of $\sim5 (\sim 57)$ Myr. This ensures that the temporal separation between two consecutive snapshots is less than the typical dynamical times of the haloes; this  is important for accurate halo-finding and merger-tree constructions, especially for merging and accreting haloes, as is common at high redshifts.
\item There are 73 snapshots between $z=5$ and $0$.  This cadence is tailored to accurately track the orbits of individual particles at late times (see \S~\ref{ssec:pericentre}). \resp{These snapshots are evenly spaced in time with a difference of $\sim175$ Myr \citep[also see][]{walkerThreeHundredProject2025}.
}
\end{itemize}

\subsection{Milky Way analogue selection}
\label{ssec:halo_selection}
We wish to study haloes whose properties are broadly consistent with those we might expect for the Milky Way's host halo at $z=0$. To obtain these Milky Way Analogues (MWAs), we enforce a \emph{mass criterion} and a \emph{neighbourhood criterion}. Our mass criterion targets haloes with masses in the range $1.04\times10^{12}~\oneh\Msun \leq {M_{\rm h}} \leq 1.95\times10^{12}~\oneh\Msun$ \citep[see][etc]{Penarrubia2015, Callingham2019}, which yields a sample of 1736 haloes.  Our neighbourhood criterion selects isolated and paired MWAs.
\begin{itemize}
\item \resp{For this mass-selected sample of 1,736 haloes, we compute the radial distance to all haloes in the simulation volume and apply the isolation criterion by requiring that no other halo more massive than the target lies within $r_{\rm iso,cut} = 1.4~\oneh \rm{Mpc}$}. There are 17 such haloes in \D{00}, which we visually inspect; from these, we choose \D{01--15}.
\item For our paired MWAs (\D{16--20}), we first select haloes that have one equally massive neighbour within $1~\oneh\rm{Mpc}$ from the mass selected sample of 1736 haloes, and we then retain only those pairs that have no haloes within $r_{\rm iso,cut} = 0.5~\oneh\rm{Mpc}$. This yields 21 systems, from which we select 5 pairs.
\end{itemize}

\begin{figure*}
\centering
\includegraphics[width=1.01\textwidth]{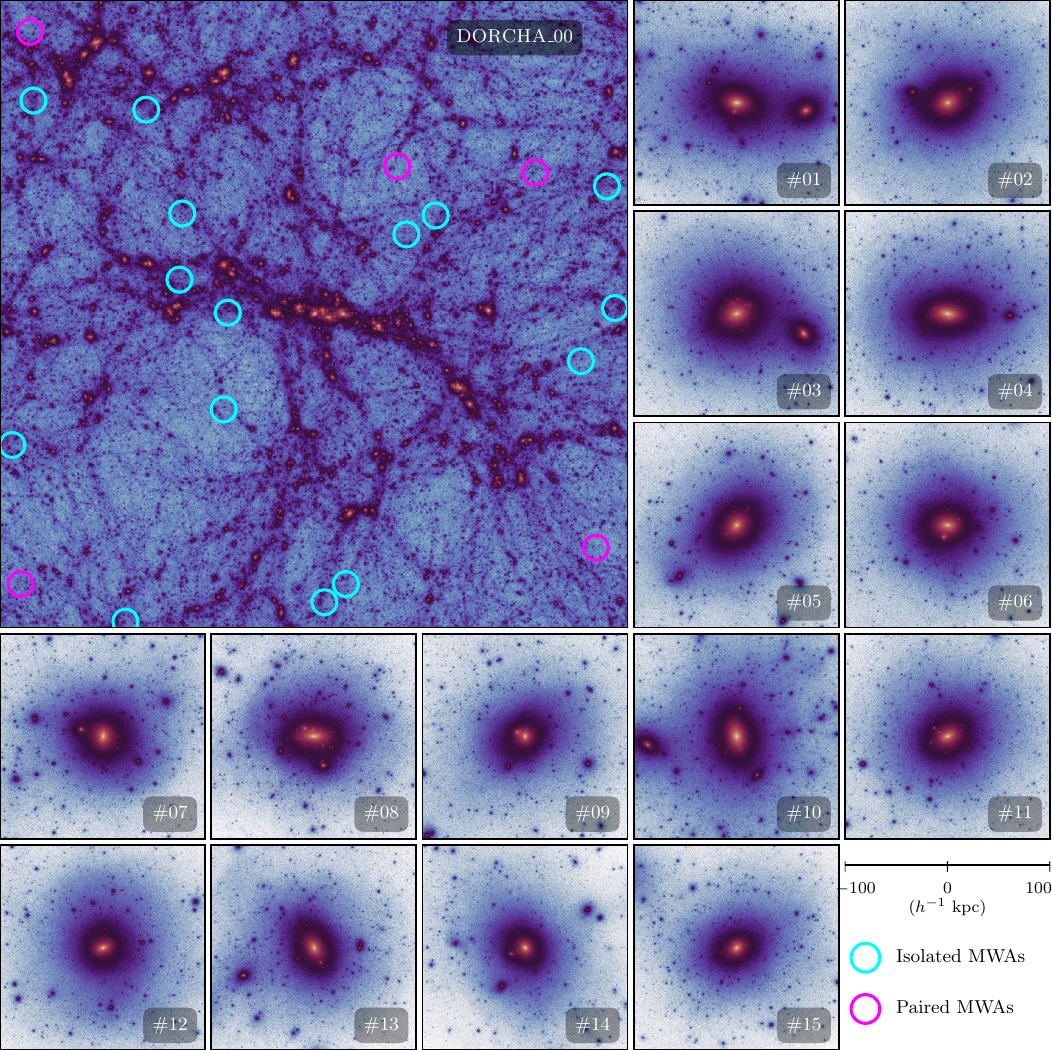}
\caption{ \textbf{Isolated \dorcha\,haloes along with the parent \D{00}}: The subplots show the total projection along the $z$-direction for our simulations. The top-left larger subplot shows $100~\oneh{\rm Mpc}\times100~\oneh{\rm Mpc}$ projection of our parent \D{00} simulation, with the locations of the isolated and paired Milky Way Analogues (MWAs) - our \dorcha~haloes - indicated by the blue and pink circles respectively. The smaller subplots show $200~\oneh{\rm kpc}\times200~\oneh{\rm kpc}$ cutouts around the \dorcha~haloes from the \D{01-15} simulations. These figures illustrate the diverse nature, in number as well as spatial distribution, of the subhaloes within the \dorcha~suite.}
\label{fig:parent1}
\end{figure*}

\begin{figure*}
\centering
\includegraphics[width=1.01\textwidth]{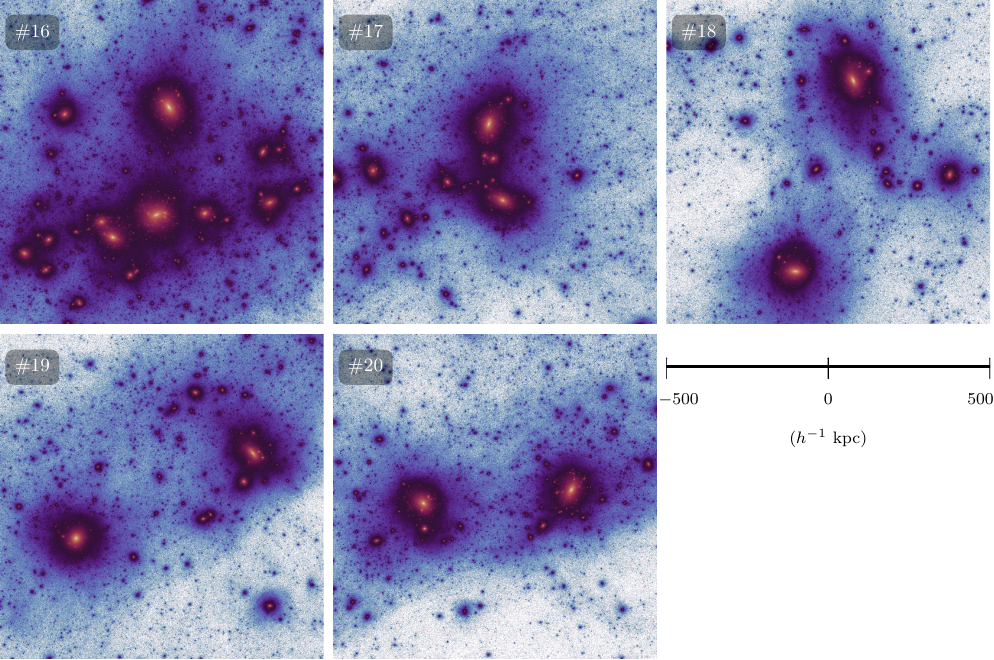}
\vspace{-0.2cm}
\caption{ \textbf{Paired \dorcha\,haloes}: Similar to Figure \ref{fig:parent1}, but for the paired \dorcha~simulations (\D{15-20}). The cutouts show $1~\oneh{\rm Mpc}\times1~\oneh{\rm Mpc}$ centred at the centre-of-mass of the two haloes in the pair.}
\label{fig:parent2}
\end{figure*}

Figures~\ref{fig:parent1} and \ref{fig:parent2} provide a visual impression of the parent simulation and the immediate environments of the isolated and pair MWAs respectively. Here we project the mass density along the $z$-direction onto the $x$-$y$ plane, such that lighter(darker) hue indicate higher(lower) dark matter density. The large top-left subplot of Figure~\ref{fig:parent1} shows the entire (i.e. $100~\oneh{\rm Mpc} \times 100~\oneh{\rm Mpc}$)  \D{00} (parent) volume, with blue and pink circles indicating the locations of the isolated and paired MWAs; the remaining subplots show regions of $200~\oneh{\rm kpc} \times 200~\oneh{\rm kpc}$ centred on the isolated MWAs,  \dorcha s (\#01 to \#15). Figure \ref{fig:parent2} shows regions of $1~\oneh{\rm Mpc} \times 1~\oneh{\rm Mpc}$ centred on the centre-of-mass of the paired MWAs, \dorcha s (\#16 to \#20).

\smallskip

\begin{figure}
    \centering
    \includegraphics[width=\columnwidth]{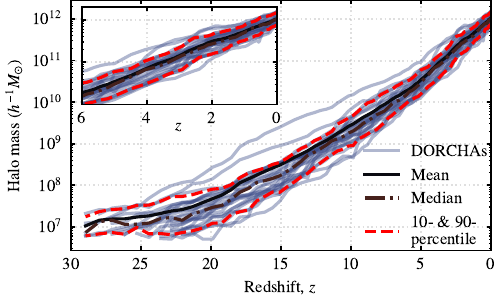}
    \caption{\label{fig:merger_history}
    \textbf{Halo growth histories}: We show the mean(black), median(brown), 10-and 90-percentile(red), and individual(light-grey) growth histories of all 25 \dorcha~haloes.  The inset on the top-left zooms in onto $z\leq6$. \textit{Takeaway:} Almost all of the haloes have progenitors that are resolved by $z\sim25$ and have formation histories that are relatively smooth.}
    
\end{figure}

\begin{figure}
\includegraphics[width=\columnwidth]{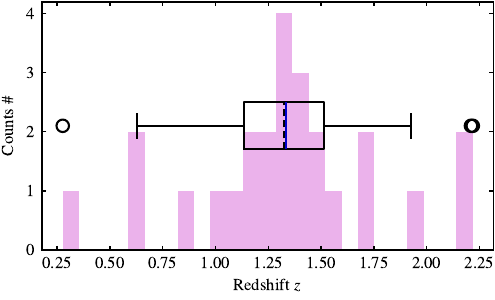}
    \caption{\textbf{Formation redshift, $z_{\rm form}\equiv z_{50}$}: Distribution of the redshifts at which 50 per cent of the present day mass of the \dorcha~haloes were \resp{assembled}. See also the last column of Table \ref{tab:sim_info}. The box-plot encompasses the 1-$\sigma$ values. Indicated also are the mean (=1.325) in solid and median (=1.33) in dashed vertical lines. \textit{Takeaway:} The \dorcha~haloes exhibit a variety of formation redshifts, enabling us to cleanly separate out the impact of the surroundings (early- and late-forming kinds) and mergers.}
    \label{fig:zform}
\end{figure}

In Figures~\ref{fig:merger_history} and \ref{fig:zform} we characterise the mass assembly histories for our sample of \dorcha s. In Figure~\ref{fig:merger_history}, the black(brown) curve shows the mean(median) mass growth, while the red dashed curves bracket the interdecile range (i.e. $10^\text{th}$ to $90^\text{th}$ percentiles); this indicates that our haloes have relatively quiescent assembly histories, which is a consequence of our choice of neighbourhood criterion that means the \dorcha s are relatively isolated systems. In Figure~\ref{fig:zform}, we show the distribution of the \dorcha s' formation redshifts, $z_{\rm form}$; the mean(median) of the distribution is 1.325(1.33). This reveals that the \dorcha s have a wide variety of assembly histories, with some forming as early as $z_{\rm form}=2.22$ (\D{02}) and others (\D{01}) forming as late as $z_{\rm form} = 0.28$.

\smallskip

Table \ref{tab:sim_info} lists properties of all 25 \dorcha~haloes. Note that among the paired \dorcha s, we follow the naming convention such that the more(less) massive one is labelled \textsc{A}(\textsc{B}). Apart from standard summary statistics to characterise the haloes' virial properties at $z$=0 ($M_{200}$, $R_{200}$, $V_{200}=(GM_{200}/R_{200})^{1/2}$), we also provide information about the number of subhaloes associated with them ($N_{\rm subs}$); when they was first identified in the \textsc{SubFind-HBT} catalogues ($z_\text{init}$); and when 10 and 50 per cent of their $z$=0 mass was first in place ($z_{10}$ and $z_{50}$). \resp{We note again that we require a minimum of 32 particles during halo-finding, yielding a subhalo mass resolution of $5.04\times 10^6~\oneh\Msun$.} The "kind" keyword indicates whether a halo is considered to be early or late forming.

\subsection{Splitting by formation time}
\label{ssec:zform}
To better understand the impact and explore if there is any correlation between the particles and the formation time of the haloes, we further categorize the \dorcha{} haloes into late- and early-forming ones. As mentioned earlier,  the formation redshift is defined as the redshift at which 50 per cent of the halo mass at $z=0$ has been formed. We use this $z_{\rm form}$ (second-to-last column of Table \ref{tab:sim_info} and Figure \ref{fig:zform}) as proxy for the amount of mergers/accretion that the halo has gone through: the halo has a higher chance of having undergone mergers before its $z_{\rm form}$. 

The mean(median) formation redshift for our haloes is 1.325(1.33). To create our lists of late- and early-\dorcha s, we select 5 haloes each furthest from above and below the mean $z_{\rm form}$. Thus selected, the late-forming haloes are \D{01} ($z_{\rm form}=0.28$), \D{18A} ($z_{\rm form}=0.63$), \D{19B} ($z_{\rm form}=0.66$), \D{17A} ($z_{\rm form}=0.89$), and \D{20A} ($z_{\rm form}=1.0$). The early-forming haloes are \D{02} ($z_{\rm form}=2.22$), \D{19A} ($z_{\rm form}=2.21$), \D{06} ($z_{\rm form}=1.93$), \D{10} ($z_{\rm form}=1.72$), and \D{05} ($z_{\rm form}=1.69$). 

\subsection{Tagging High Redshift Remnants}
\label{ssec:highz_sf}

\resp{We aim to identify regions of phase space that are likely to host remnants of star formation originating in the high-redshift progenitors of Milky Way analogues that later reside within our \dorcha~haloes.}
For each \dorcha~halo, we identify and tag the particles that are associated with its progenitor haloes at redshifts $z\geq5$. We do this by creating lists of tagged particles at $z\in[25, 20, 15, 10, 8, 5]$. Where a halo's progenitors are not resolved by $z_{\rm init}=25$, we start tagging at the first redshift at which a progenitor is recovered (sixth column of Table~\ref{tab:sim_info}). At each of these redshifts, we tag the most-bound 10 per cent\footnote{We have repeated our analyses with 5 and 20 per cent cuts. The precise value of this per cent threshold does not impact our conclusions.}  of particles in each of the progenitor haloes -- this ensures that we follow the cores of these progenitors. The minimum halo mass of the \dorcha~suite -- $\mathrm{M}_\text{h,min}\simeq5.04\times10^6~\oneh\Msun$ -- is comparable to the halo mass at which cooling becomes efficient at $z\sim40-20$ \citep[cf.][]{Nebrin2003}, and so we might expect these cores to trace where star formation is likely to occur. In this way, we can identify the material in a $z=0$ descendant that was once in the cores of its progenitors at $z\geq5$, which could trace where we might expect to find the oldest stars and their remnants.    

For comparison, we also tag particles based on when they first formed part of a halo -- "old" material was in place in progenitor haloes before $z=10$, while "new" material was accreted onto the main branch of the halo after $z=5$. 

\help{We remind the reader that the \dorcha~suite is, by construction, dark-matter-only, and so we caution that our predictions miss the potential impact of baryon physics, even though our prescription for assigning the sites of star formation are rooted in physically sound considerations. Having said this, the conclusions we draw from the \dorcha~suite will serve as a baseline to compare against for our future simulations, which will explicitly model hydrodynamical galaxy formation. 
We emphasise that while our tagged particles are intended to trace the sites of early star formation, we do not explicitly distinguish between long-lived Population III stars and their extremely metal-poor (but not metal-free) descendants, although these populations may have different formation histories and survival rates. See $\S$\ref{ssec:remnants_today} for relevant discussion.}

\subsection{Measurement of Radial Profiles}
\label{ssec:spherical_averaging}
Radial profiles can be affected by the particular choice of the centre of the halo \citep[cf.][]{navarroDiversitySimilaritySimulated2010}. When constructing spherically averaged profiles, we select particles within $r/R_{200}\leq 3$ of the halo centre, which we define as the position of the minimum gravitational potential  recovered by \textsc{SubFind-HBT}. These particles are then binned into 50 radial bins spaced evenly in logarithm between $10^{-3}\leq r/R_{200}\leq 3$  for the radial profiles. The radius of each shell is computed from the mean radii of the particles within the shell. 

The mass density profile ($\rho(r)$) is obtained by summing over the mass of particles in the shell and dividing by the shell volume. The mean radial velocity ($\overline{v}_r$) and velocity dispersion ($\sigma_r$) profiles are calculated from
\begin{equation}
\overline{v}_r=N_{\rm shell}^{-1}\sum_i\frac{\vec{r}_i}{r_i}\cdot(\vec{v}_i-\vec{v}_{\rm cm}),
\end{equation}
which gives the mean radial velocity within a shell, with $\vec{r}_i$ being the position vector of the $i$th particle and $\vec{v}_{\rm cm}$ the centre-of-mass velocity of the halo;
\begin{equation}
\overline{v_r^2}=N_{\rm shell}^{-1}\sum_i\left[\frac{\vec{r}_i}{r_i}\cdot(\vec{v}_i-\vec{v}_{\rm cm})\right]^2,
\end{equation}
which gives the mean square radial velocity within a shell; and 
\begin{equation}
\sigma_r=\left[\overline{v_r^2}-\overline{v}_r^2\right]^{1/2},
\end{equation}
which gives the corresponding radial velocity dispersion in the shell. We compute the total velocity dispersion as,
\begin{equation}
\sigma^2_{\rm 3D}=\sigma^2_{\rm x}+\sigma^2_{\rm y}+\sigma^2_{\rm z}=\sigma^2_{\rm r}+\sigma^2_{\rm tan},
\end{equation}
which allow us to obtain the tangential velocity dispersion $\sigma_{\rm tan}$ from the radial ($\sigma_{\rm r}$) and Cartesian components of the velocity dispersion ($\sigma_{\rm x/y/z}$). This allows us to compute the velocity anisotropy,
\begin{equation}
\beta(r)=1-\frac{1}{2}\frac{\sigma^2_{\rm tan}}{\sigma^2_{\rm r}},
\end{equation}
which characterises how radial orbits are; for example, $\beta\rightarrow1$ indicates preferentially radial orbits, while $\beta\rightarrow0$ will be preferentially isotropic. In addition, we estimate the mean rotational velocity within each radial shell as,
\begin{equation}
   \overline{v}_{\rm rot}=J_{\rm shell}M^{-1}_{\rm shell}R^{-1}_{\rm shell}
\end{equation}
where $J_{\rm shell}\equiv\left|\vec{J}_{\rm shell}\right|$ is the magnitude of the angular momentum vector within the shell relative to the centre of mass velocity of the halo, and $M_{\rm shell}$ and $R_{\rm shell}$ are the mass and radius of the shell respectively\footnote{\resp{We caution that the angular momentum orientation may vary with radius. Consequently, adjacent shells can be dominated by different orbital streams with similar $\left|J_{\rm shell}\right|$ but differing orientations, leading to comparable $\overline{v}_{\rm rot}$ despite distinct underlying kinematic structures.}}

\section{Results}
\label{sec:results}
In the following subsections, for our sample of $z=0$ \dorcha~haloes, we examine the spherically averaged profiles, assessing their mass density and kinematic profiles (\S~\ref{ssec:profiles}). We also study the phase space distributions (\S~\ref{ssec:phase_space}) and the orbital distributions and pericentres (\S~\ref{ssec:pericentre}) of material within $R_{200}$. 
\resp{Our aim is to isolate regions in phase space where we are most likely to find material originating from the high-redshift progenitors of Milky Way analogues, and to assess the extent to which these results depend on a MWA’s assembly history and the dynamical evolution of its central regions.
}

\subsection{Spherically Averaged Profiles of High Redshift Remnants}
\label{ssec:profiles}
\begin{figure*}
    \centering
    \includegraphics[width=1.01\textwidth]{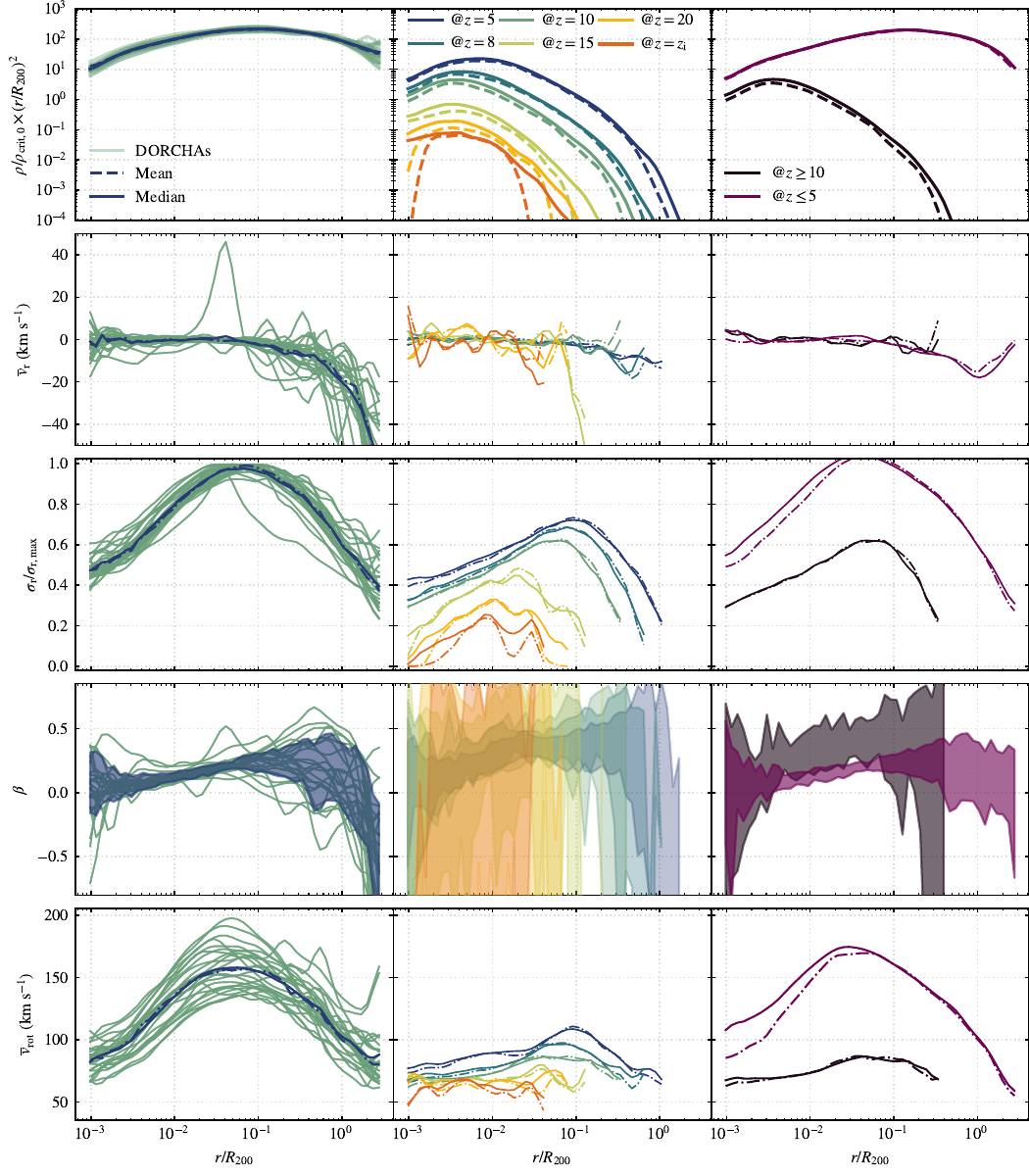}
    \caption{\label{fig:radial_profiles}\textbf{Radial profiles for all \dorcha s}: The rows from top to bottoms shows, respectively, the density profile $(\rho)$, mean radial velocity $(\overline{v}_{\rm r})$, radial velocity dispersion $(\sigma_{\rm r})$, velocity anisotropy $(\beta)$, and mean rotational velocity $(\overline{v}_{\rm rot})$. From left to right the columns are for, respectively, all the particles that comprise the halo, the particles that were tagged at high redshifts in different colours, and the particle list for old $(z\geq10)$ and new $(z\leq5)$ particles. Mean(median) trend is shown with solid(dashed) curves. The shaded region for the $\beta$-profiles show the 1-$\sigma$ variation. The particles are binned into 50 radial bins evenly spaced in logarithm between $10^{-3}\leq r/R_{200}\leq 3$. \textit{Takeaway}: The radial profiles in the 2$^{\rm nd}$-column shows that most of the high-$z$ progenitors are now part of the main subhalo at $z=0$. The earlier a particular set of particles were tagged, the more centrally concentrated they are at $z=0$.}    
\end{figure*}
In Figures \ref{fig:radial_profiles}, \ref{fig:radial_profiles_early}, and \ref{fig:radial_profiles_late} we show spherically averaged radial profiles for the \dorcha~haloes. In each figure, we show (from top to bottom), 
\begin{enumerate}
    \item  the radial mass density profile in the form $\rho(r)\times\,r^2$, normalised by $\rho_{\rm crit}\times\,R_{200}^2$; 
   \item the radial velocity, $\overline{v}_{\rm r}(r)$, in units of km s$^{-1}$; 
   \item the radial velocity dispersion, $\sigma_{\rm r}(r)$, normalised by its peak value, $\sigma_{\rm r,max}$;
   \item the velocity anisotropy, $\beta(r)$; and 
   \item the mean rotational velocity, $\overline{v}_{\rm rot}$, in units of km s$^{-1}$. 
\end{enumerate}
The left column shows the results for all of the material within each of the haloes at $z$=0; the middle column focuses on only those particles that were in the high redshift ($z\geq5$) progenitors of the $z$=0 halo, and shows profiles for $5\leq z \leq z_i=25$; and the right column shows the results for material that were in place at $z\geq10$ ("old") and material accreted at $z\leq5$ ("new"). In the first column we show the curves for all the \dorcha s (in light-green) while for the other columns we only show the average trends -- mean(median) trend in solid(dotted) curves. 

\subsubsection{The full \dorcha~sample}
\noindent\emph{Mass Density:} The top left panel of Figure \ref{fig:radial_profiles} shows that the mass density profiles of each of the 25 \dorcha~haloes (green curves) exhibit little variation with respect to each other out to $R_{200}$, irrespective of whether they are isolated or paired; consistent with \cite{garrison-kimmelELVISExploringLocal2014}. Relative to the median profile (dark solid curve), which overlaps the mean profile (dark dashed curve), we find that the measured $\rho(r)$ differs by approximately no more than 30 per cent within $\sim R_{200}$, and the most significant deviations arise in the outskirts.

In the top middle panel, we focus on those particles that are associated with high redshift progenitors. The 6 curves correspond to the present day median (solid) and mean (dashed) profiles of particles that were already in progenitors that formed by $z=(5,8,10,15,20,z_i)$. As we expect, the density profiles of material associated with earlier forming progenitors have a lower amplitude (because they contain less mass); they are more centrally concentrated (the radius, $r_{-2}$, at which the peak of the $\rho\,r^2$ profile occurs increases by approximately 0.5 dex for material associated with $z=z_i$ and $z=5$ progenitors); and they truncate at progressively smaller radii. Interestingly, these profiles have a similar shape, with an outer slope that can be approximated as $\rho\propto\,r^{-4}$.

In the top right panel, we separate particles into those that were in progenitors prior to $z=10$ (the old component) and those accreted since $z=5$ (the new component); as before, the mean and medians are shown by the dashed and solid curves. This emphasises how much more centrally concentrated the old component is relative to the young component ($r^{\rm old}_{-2}/R_{200}\simeq0.003$ compared to $r^{\rm new}_{-2}/R_{200}\simeq0.1$) and how much more rapidly the mass density of the old component drops off relative to the new ($\rho^{\rm old}\,\sim\,10^{-1}\,\rho^{\rm new}$ at 0.003$R_{200}$ to $\rho^{\rm old}\,\sim\,10^{-3}\,\rho^{\rm new}$ at 0.1$R_{200}$).

\smallskip

\noindent\emph{Kinematics:} The middle three rows of Figure \ref{fig:radial_profiles} show the mean radial velocity ($\overline{v}_r$); radial velocity dispersion ($\sigma_r$); and velocity anisotropy ($\beta$) profiles. 

In the left panels, we show the kinematic profiles for each of the 25 \dorcha~haloes in green with the median and mean profiles overlaid by the heavy navy solid and dashed curves. The median behaviour is for $\overline{v}_r\simeq0$ within $\sim0.8\,R_{200}$, before declining in the outskirts, at $\sim\,R_{200}$, as a result of infalling material. Individual haloes show local variations of $\pm 20~{\rm km ~s}^{-1}$ in the outskirts, while one (\D{17B}) has a nearby companion which manifests as a spike at $10^{-2}\leq r/R_{200} \leq 10^{-1}$. The median velocity dispersion peaks at $\simeq 0.06\,R_{200}$; individual profiles typically fall within $\pm 5\%$, with the exception of \D{17B}. The velocity anisotropy increases from $\beta\simeq 0$ (isotropic) in the centre to $\beta\simeq0.3$ (preferentially radial) at $R_{200}$. The shaded region indicates the 10-to-90 percentile range, which shows that individual haloes have velocity distribution in the outskirts that vary from isotropic ($\beta\simeq0$) to strongly radial ($\beta\simeq0.5$).

In the middle panels, we plot the median and mean kinematic profiles for the material associated with the high redshift progenitors identified between $z=5$ and $z=z_i$. The median behaviour is for $\overline{v}_r\simeq0$, which indicates that the material is in dynamical equilibrium\footnote{\resp{We caution the reader that a vanishing $\overline{v}_r$ is a necessary but not sufficient condition for dynamical equilibrium.}}; there is no evidence for a decline in $\overline{v}_r\simeq0$ in the outskirts, which implies that all of the progenitor material has been assembled in the main body of the halo for several dynamical times. The median velocity dispersion of the $z=5$ component peaks at $\simeq 0.06\,R_{200}$; while its shape is preserved, the magnitude and location decreases as we probe material from higher redshift components \citep[c.f.][]{navarroDiversitySimilaritySimulated2010}. This indicates that the material that assembled earliest in progenitors that collapsed at $z\geq 10$, which is most centrally concentrated, is dynamically colder -- as we would expect for material in the innermost parts of the gravitational potential. The velocity anisotropy profiles for each of the components have similar shapes, showing a characteristic increases from $\beta\simeq 0$ (isotropic) in the centre to $\beta\simeq0.5$ (preferentially radial) at $R_{200}$ \citep[see also][]{loebmanBetaDipsGaia2018, rileyVelocityAnisotropyMilky2019}. As the shaded regions reveal, there is a larger variation in $\beta$ but the general trend is for more strongly radial orbits for the higher redshift components. \resp{For instance, for the $r/R_{200} = 10^{-2}$ bin, the median(scatter) of $\beta$~for $z \in [15, 10, 5]$ is, respectively,  0.43(0.3479), 0.4286(0.2284), 0.241(0.1928).}

In the right panels, we plot the median and mean kinematic profiles for the material associated with the old and new components ($z\geq10$ and $z\leq 5$ respectively). The trends are similar to those in the middle panels; the median $\overline{v}_r\simeq0$, which indicates that the material is in dynamical equilibrium, while the median velocity dispersions of the two component peaks at $\simeq 0.06\,R_{200}$, differing only in magnitude. The inner slope of the velocity dispersion profile of the old component is shallower than that of the new component, as we would expect for material orbiting predominantly in the innermost parts of the potential for many dynamical times (see \S~\ref{ssec:pericentre} for more discussion). 
\resp{This is reflected in the differences between the velocity anisotropy profiles; while both increase with radius, the older component is more strongly radially biased. The turnover at $\sim0.1R_{200}$ is consistent with an apocentre pile-up of material on eccentric orbits, though this does not by itself imply tangential anisotropy, since velocity anisotropy reflects the relative dispersions of the velocity components rather than the mean orbital phase.}

\smallskip

\noindent\emph{Angular Momentum:} In the bottom row of Figure \ref{fig:radial_profiles}, we show the mean rotational velocity, $\overline{v}_{\rm{rot}}$. The trend shows the expected one for the dark matter haloes. The $\overline{v}_{\rm{rot}}$ rises with radius in the inner regions of the halo due to the steep mass profile. Further away from the centre, the $\overline{v}_{\rm{rot}}$ declines as the density decreases. We note that, from the third column, the older material have a comparatively flat $\overline{v}_{\rm{rot}}$ profile. 

\smallskip

\help{
\noindent\textbf{General comments:} 
There are few markedly different features between the old and the new components that could help one distinguish between them observationally. If, within a given radial bin, we can identify the subset of stars with low radial and rotational velocities when compared to the full population in that bin, then there is a high chance that they belonged to one of the high-$z$ ($z\geq10$) progenitors of the MW. Our results are in qualitative agreement with \citep{arentsenPristineInnerGalaxy2020}, who studied the kinematics of the metal-poor stars in the Galactic bulge as a function of metallicity and reported a decreasing rotational velocity with decreasing metallicity for the metal-poor stars. This is interesting because it suggests that the impact of galaxy assembly and the presence of a stellar bar on the inner potential will affect our results quantitatively but not qualitatively. Of course, we can verify this in our forthcoming simulations that model hydrodynamical galaxy formation.}

\subsubsection{The early- and late-forming \dorcha~sample}

\begin{figure*}
    \includegraphics[width=1.01\textwidth]{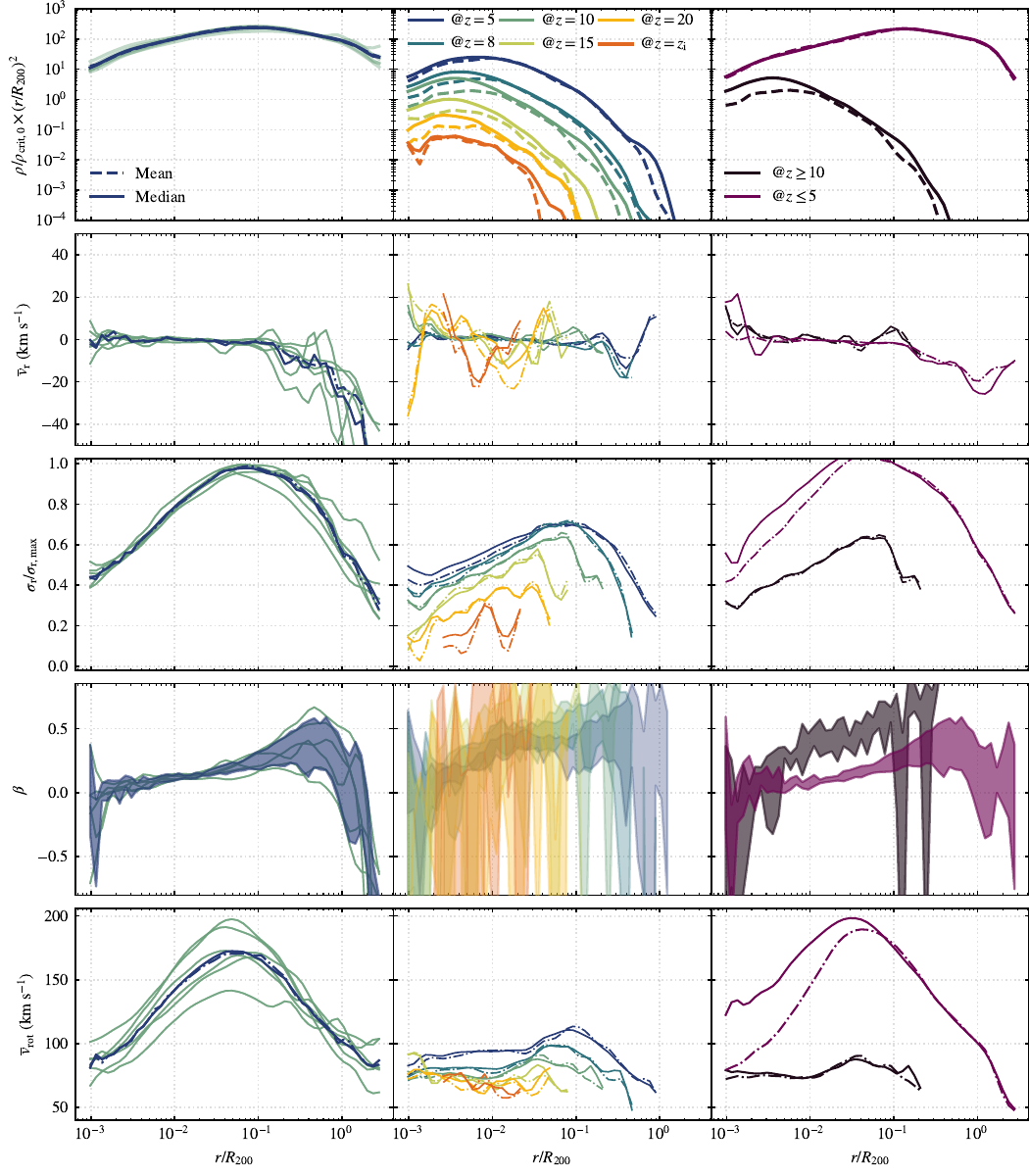}
     \caption{\label{fig:radial_profiles_early}\textbf{Radial profiles for the early-\dorcha s}: Similar to Figure \ref{fig:radial_profiles} but only for 5 of the earliest-forming \dorcha s ("early" kind). These are \D{02} ($z_{\rm form}=2.22$), \D{19A} ($z_{\rm form}=2.21$), \D{06} ($z_{\rm form}=1.93$), \D{10} ($z_{\rm form}=1.72$), and \D{05} ($z_{\rm form}=1.69$). Note that all of them, except for \D{19A} are isolated \dorcha s.
     \textit{Takeaway}: The mean and median trends are very similar to the full set of \dorcha s (see Figure \ref{fig:radial_profiles}).
    }
\end{figure*}
\begin{figure*}
    \includegraphics[width=1.01\textwidth]{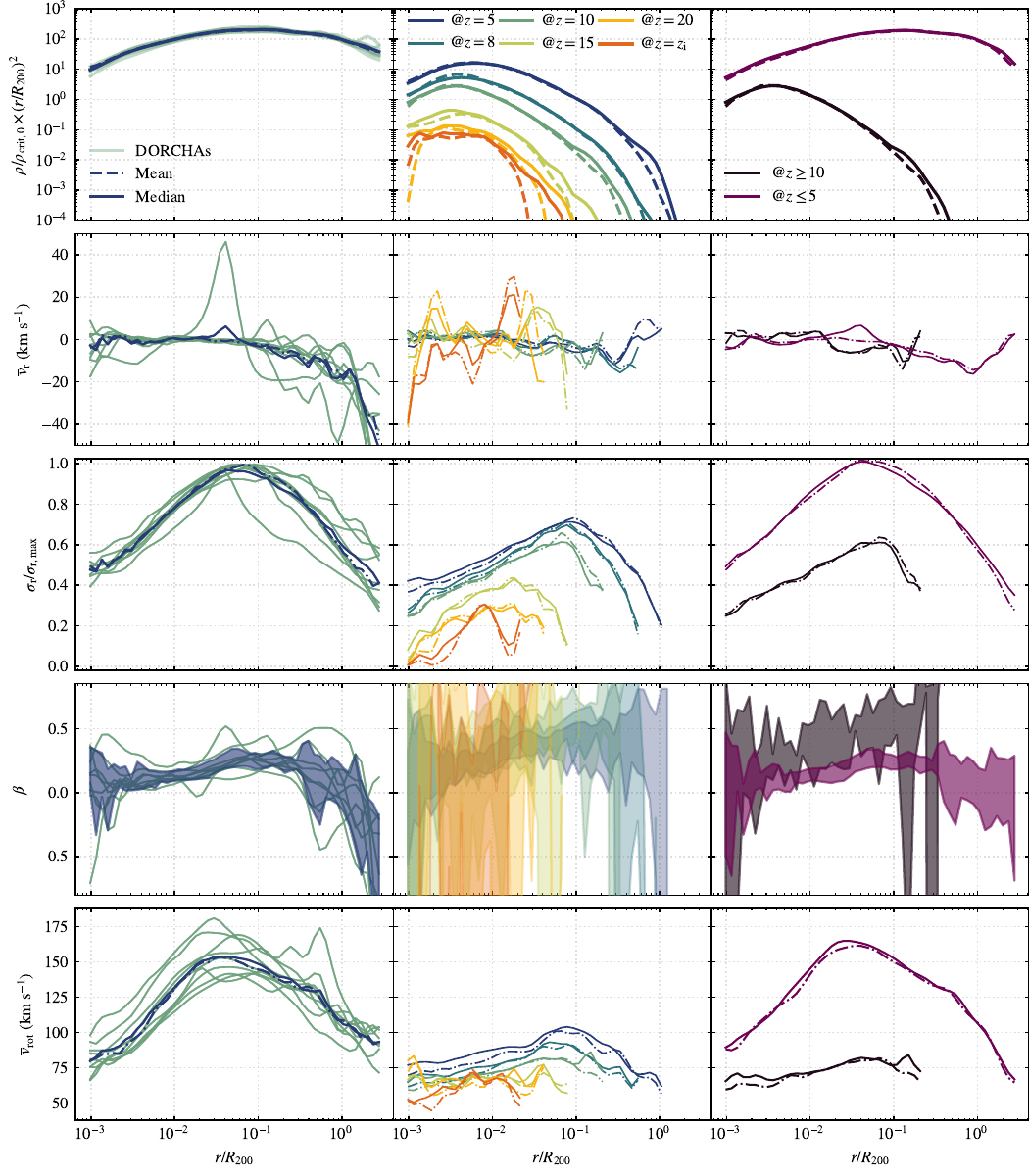}
     \caption{\label{fig:radial_profiles_late}\textbf{Radial profiles for the late-\dorcha s}: Simialr to Figures \ref{fig:radial_profiles} \& \ref{fig:radial_profiles_early}, but only for 5 of the late-forming \dorcha s. These are \D{01} ($z_{\rm form}=0.28$), \D{18A} ($z_{\rm form}=0.63$), \D{19B} ($z_{\rm form}=0.66$), \D{17A} ($z_{\rm form}=0.89$), and \D{20A} ($z_{\rm form}=1.0$), almost all of whcich are paired \dorcha s. \textit{Takeaway}: The primary difference difference cf. Figure \ref{fig:radial_profiles_early} is the $\beta(r)$ profile. Any other minute differences in kinematics can be explained by the presence of a nearby massive neighbour.}
     
\end{figure*}

In Figure~\ref{fig:merger_history} we showed that the \dorcha~haloes have relatively quiescent histories with no evidence for major mergers in their recent assembly history. However, for each halo we have a formation redshift, $z_{\rm form}$ -- the redshift at which half of the mass of the main progenitor of the $z$=0 halo is first in place (see \S~\ref{ssec:zform}) -- which we show in Figure~\ref{fig:zform}. Based on this information, we select the five earliest and five latest forming haloes  (i.e. those with the largest and smallest  $z_{\rm form}$ values respectively) from the \dorcha~sample and repeat our analysis of the previous subsection in Figures \ref{fig:radial_profiles_early} and \ref{fig:radial_profiles_late}.

Despite the difference in the typical formation redshift of the earliest-forming haloes ($z_{\rm form}\simeq2$)  compared to the latest-forming haloes ($z_{\rm form}\simeq0.6$), we find remarkably little variation in the mean and median profiles. This is surprising because the latest-forming haloes are preferentially those in pairs, where we might expect the presence of a massive neighbour to have an influence. The mass density profiles are consistent with one another -- the trends we noted in the previous section remain regardless of how we split by $z_{\rm form}$ -- while differences in the kinematics profiles can be explained as arising from the proximity of a massive neighbour rather than differences in $z_{\rm form}$. 

The main difference of note is in the velocity anisotropy -- if we consider the left-most panels, we see that the latest-forming \dorcha~haloes are mildly preferentially radial ($\beta\lesssim 0.2$) at $r\gtrsim 0.5R_{200}$, whereas the earliest-forming ones are more strongly preferentially radial ($\beta\lesssim 0.5$) at $r\simeq R_{200}$. However, this reflects environmental differences and the influence of a nearby massive halo.  

Note that if we select haloes by $z_{10}$ rather than $z_{\rm form}\equiv\,z_{50}$, we find that the results are unchanged -- this reflects the relatively quiescent assembly histories of the haloes in our sample, such that rank ordering of the distribution is similar for either $z_{10}$ or $z_{50}$. 

\smallskip

\subsection{Phase Space Structure of High Redshift Remnants}
\label{ssec:phase_space}

\begin{figure*}
    \centering
    \includegraphics[width=1.01\textwidth]{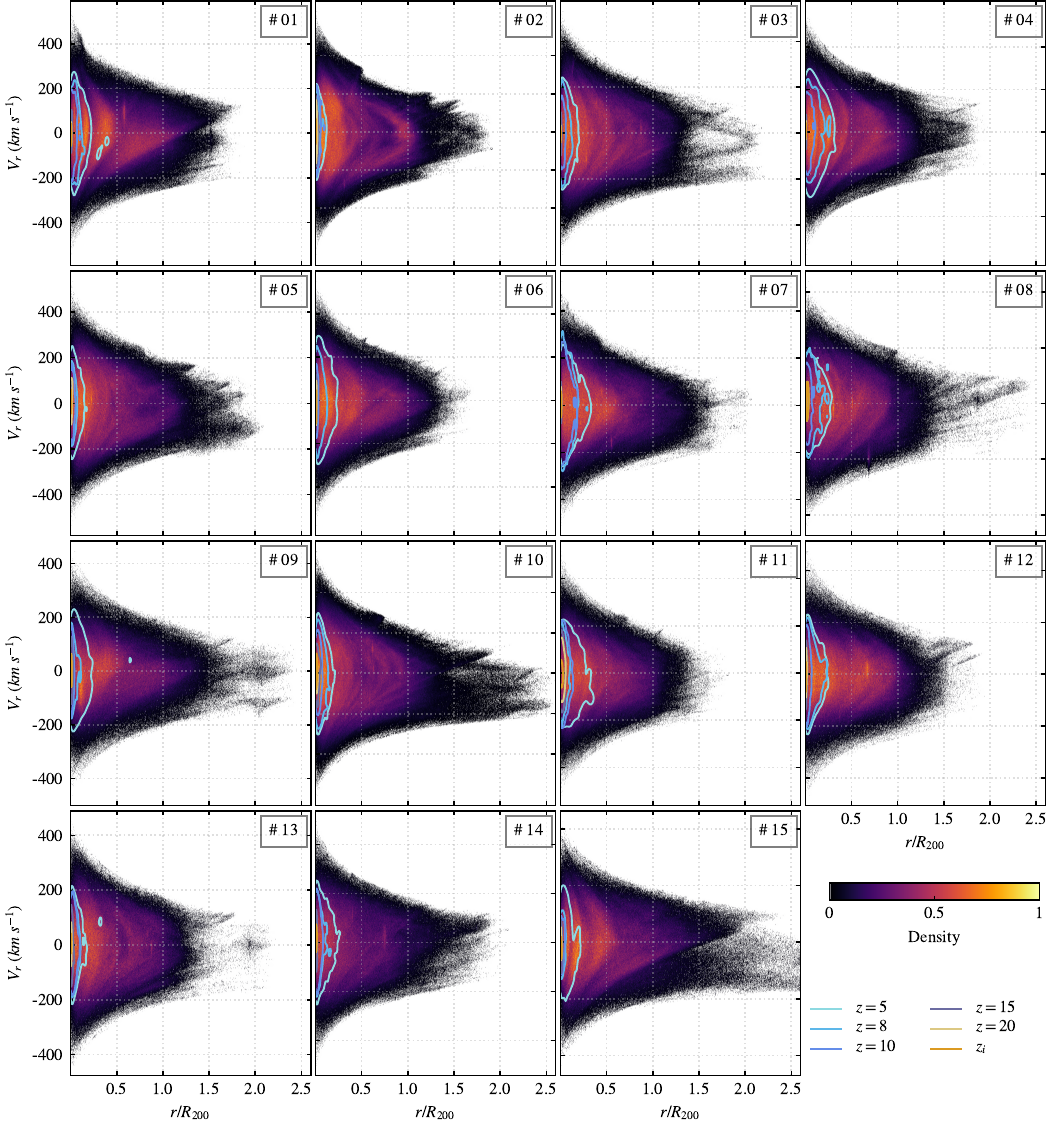}
    \caption{\label{fig:phase_plot_iso}\textbf{Phase plot for the isolated \dorcha s:}: Shown are the phase-space distribution, i.e. the radial velocity $(v_{\rm r})$ against the radial distance $(r)$), for the entire halo. The lines represent the 90 per cent contour lines of the distribution for particles that were tagged at the indicated redshifts. \textit{Takeaway:} Multiple caustics can be clearly seen in many of the \dorcha s. It is evident that most of the tagged high-$z$ progenitor sites have sunk to the centre of the of the halo (see also Figure \ref{fig:phase_plot_iso_contours}).}
\end{figure*}

\begin{figure*}
    \centering
    \includegraphics[width=1.01\textwidth]{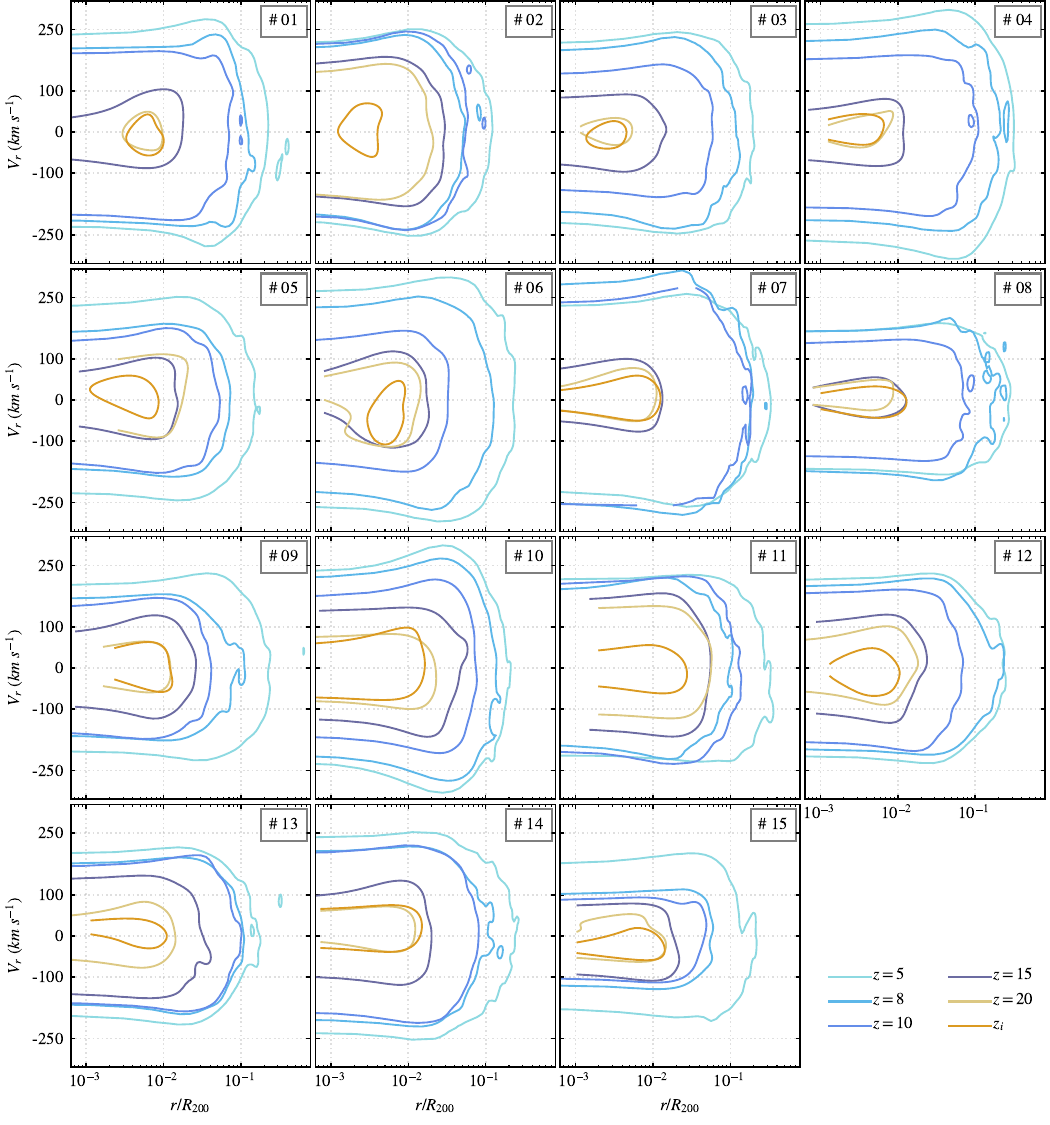}
    \caption{\label{fig:phase_plot_iso_contours}\textbf{Phase plot for the isolated \dorcha s}: A zoom in on the contours shown in Figure \ref{fig:phase_plot_iso}. \textit{Takeaway:} The bulk of the material associated with the high-$z$ progenitors is confined to the innermost parts of the potential, and is dynamically mixed,  relaxed and cold.}
\end{figure*}

The results of the previous section have highlighted that material associated with haloes that collapsed at $z\geq5$ is 
\begin{enumerate}
    \item concentrated in the innermost parts of the halo today, with a density profile that declines as $r^{-4}$ with increasing radius; 
    \item in dynamical equilibrium (i.e. $\overline{v_r}\simeq0$); 
    \item dynamically colder, as measured by $\sigma_r$, and has a more strongly preferentially radial velocity distribution, as measured by $\beta$, than the bulk of the material that forms the halo. 
\end{enumerate}These results derive from spherically averaged radial profiles, but what kind of structures might we expect in phase space?  

\smallskip

In Figures \ref{fig:phase_plot_iso} and \ref{fig:phase_plot_iso_contours}, we show the phase space distribution of halocentric radius, $(r)$, in $h^{-1}$ kpc against radial velocity, $(v_{\rm r})$, in km s$^{-1}$ for the isolated \dorcha~haloes at $z$=0 (plots for the paired haloes can be found in the Appendix \ref{sec:appendix1}). The colour-coding in Figure~\ref{fig:phase_plot_iso} indicates the phase-space density at a particular value of ($r,v_r$). We include only material that is gravitationally bound to the 
the primary subhalo of the FoF group, and show contours enclosing 90 per cent of the material associated with high redshift, colour-coded by the redshift at which it was identified. In Figure \ref{fig:phase_plot_iso_contours}, we zoom in on these contours for greater clarity.

Figure \ref{fig:phase_plot_iso} reveals that the \dorcha~haloes contain a wealth of fine phase space structure. Multiple caustics -- broadly symmetric ridges of phase density extending from $-200 {\rm ~km ~s}^{-1}\lesssim v_r\lesssim 200 {\rm ~km ~s}^{-1}$ and most evident within the inner 0.25 $R_{200}$ (50 $h^{-1}$kpc) -- can be found in many of the haloes. For example, the caustics in \D{01} and \D{15} are sharply defined, although in most cases the caustics are less distinct. This fine phase space structure provides an integrated record of the orbits of disrupting substructure within the halo potential during its assembly history \citep{henriksenSelfSimilarRelaxationSelfGravitating1997}. At radii of order $0.5-1$ $R_{200}$ ($100-200$ $h^{-1}$kpc), the phase space structure becomes more asymmetric, and we see ridges with skewed $v_r<0$ that can be traced to spurs and bifurcations on the haloes' edges (such as in \D{01} and \D{15}). 
\resp{This may reflect the interaction of the haloes with their larger-scale environment, as well as the contribution from ongoing infall and disrupted material at large radii.
}

Figure \ref{fig:phase_plot_iso_contours} allows for closer inspection of the phase space distribution of the material associated with high redshift progenitors, and finds broad consistency across the sample. The contours are symmetric and centred on $v_{\rm r}=0 ~{\rm km~s}^{-1}$, as we would expect from the spherically averaged profiles in the previous section. The $z=5$ contours span the range from $-200 {\rm ~km ~s}^{-1}\lesssim v_r\lesssim 200 {\rm ~km ~s}^{-1}$ and extend to $\sim 0.1 R_{200}$ ($r\simeq 10-20~h^{-1}{\rm kpc}$), while the oldest material ($z\geq15$) spans  $-100 {\rm ~km ~s}^{-1}\lesssim v_r\lesssim 100 {\rm ~km ~s}^{-1}$ within $\sim 0.005 R_{200}$ ($r\simeq 1~h^{-1}{\rm kpc}$). There is some fine-scale structure evident in the contours -- for example, spurs in \D{09} and \D{15} and localised clumps in \D{08}, and are indicative of spatially coherent overdensities \resp{(They could also be substructures that have not be properly identified by the halo-finder)}. However, on the whole, these phase space distributions confirm that the bulk of the material associated with the high redshift progenitors is confined to the innermost parts of the potential; dynamically mixed and relaxed; and relatively dynamically cold. 

\subsection{Orbital Structure of High Redshift Remnants}
\label{ssec:pericentre}

\begin{figure}
    \centering
    \includegraphics[width=\columnwidth]{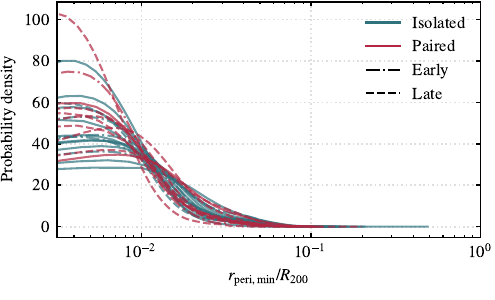}
    \caption{\textbf{Pericentric distribution}: The probability distribution of the minimum pericentric distances of the "old" (i.e. $z\geq10$) tagged particles for $z\leq5$. The pink (light green) curves denote paired (isolated) \dorcha s. The dashed (dash-dotted) represent late (early) \dorcha s. \textit{Takeaway}: For all the \dorcha s, most of the particles that we tag at high-$z$ haloes have interacted with the centre of the main subhalo. There is no  distinguishable trend that stands out for different categories of \dorcha s. This is likely a consequence of the fact that these particles have sunk into the potential well of their haloes and are thus more robust to external perturbations such as mergers and accretions.}
    \label{fig:pericentre_dist}
\end{figure}

The previous results demonstrate that material associated with high redshift remnants is centrally concentrated within the MWAs at $z=0$ -- precisely in the region where the galaxy disc will form. This suggests that the oldest remnant material should be well mixed within the inner regions of the galaxy, and its spatial and kinematic structure is likely to be modified in the flattened disc potential of the galaxy. The size of this effect will correlate with how long this material has been present within the inner regions of the halo. 

To assess this, we carry out explicit orbit tracking of all the particles associated with high redshift progenitors using the algorithm \resp{described in \cite{walkerThreeHundredProject2025}}. Our method follows particles from $z=0$ to high redshifts across consecutive simulation snapshots and logs when each one experiences a pericentric passage relative to the centre of the MWA's main progenitor; it does this by identifying a sign change in a particle's radial velocity (relative to the MWA's main progenitor's centre) from negative to positive. To filter out pericenters associated with orbits within substructure, the algorithm selects only radial velocity sign changes that occur after a particle has advanced through a minimum angle, $\theta_{\rm min}$, around the halo centre since its last sign change. We use $\theta_{\rm min}=\pi/2$.

Figure \ref{fig:pericentre_dist} shows the probability distributions (PDFs) of the minimum pericentric distances of the particles that make up the material associated with high redshift remnants, i.e. those tagged  at $z\geq10$. Here, we track their orbits from $z=5$ and determine the minimum pericentric distance of a particle during this period, as a fraction of $R_{200}$ at $z=0$.

We indicate paired and isolated \dorcha s by pink and green curves; the subset of early- and late-forming \dorcha s are indicated by dot-dashed and dashed curves respectively. As we anticipate, the distributions of minimum pericentric distances rise sharply at $r\lesssim 0.01 R_{200}$ and peak in the range $\sim 0.001-0.01R_{200}$. Combined with Figure~\ref{fig:phase_plot_iso_contours}, which suggest that the typical 90 per cent contour occurs at $\sim 0.1R_{200}$, this indicates that this material will follow high eccentricity orbits. There is no obvious distinction between haloes that are paired or isolated, or early- or late-forming, which is consistent with our previous results. 

\resp{Although here we quantify the interaction with the halo center with the `minimum pericentre' since it directly measures passage through (and mixing within) the central potential, our existing Figures \ref{fig:phase_plot_iso} -- \ref{fig:pericentre_dist} also imply a characteristic `most‑recent apocentre' for this ancient component: combining the 90\% $r - v_r$ envelopes ($\lesssim0.1\,R_{200}$) with $r_{\rm peri,min}\,\sim\,10^{-3}\,-\,10^{-2}\,R_{200}$ yields typical eccentricity values of $e\,\gtrsim\,0.8$ and apocentre values of $r_{\rm apo}\,\sim\,0.05\,-\,0.1\,R_{200}$. We note that is consistent with orbit‑based chemo‑dynamical constraints on metal‑poor stars in the inner Galaxy from COMBS and PIGS \citep[see][]{luceyCOMBSSurveyIII2021, ardern-arentsenPristineInnerGalaxy2024}}

\subsection{Looking for the remnants of high redshift star formation in the Milky Way}
\label{ssec:remnants_today}

The results we have presented have made clear that, regardless of the details of the assembly history of the MWA host halo, the material associated with the remnants of MWA high redshift progenitors will be centrally concentrated in the innermost parts of the potential at $z=0$; it will be dynamically relaxed; and the oldest components will be relatively dynamically cold compared to their surroundings. Almost all of this material associated with high redshift remnants ends up in the main subhalo of our \dorcha~haloes. \help{This is consistent with previous simulations (e.g. \citealt{tumlinsonCHEMICALEVOLUTIONHIERARCHICAL2010,gaoEarliestStarsTheir2010,starkenburgOldestMostMetalpoor2017,hortaProtogalaxyMilkyWaymass2023}; but see also \citealt{el-badryWhereAreMost2018}) and observations \citep[see e.g.][]{arentsenPristineInnerGalaxy2020, sestitoPristineInnerGalaxy2022, ardern-arentsenPristineInnerGalaxy2024}. }

\begin{figure}
    \centering
    \includegraphics[width=\columnwidth]{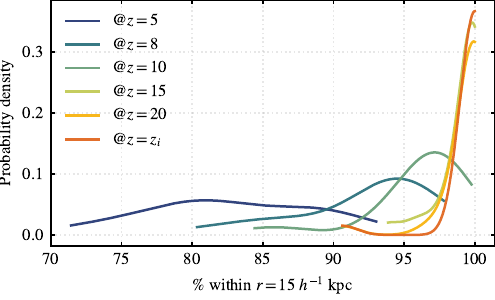}
    \caption{\textbf{Spatial distribution of tagged particles:} The probability distribution of the present-day fraction  of tagged particles that are within $r\leq15~\oneh{\rm kpc}$ of the halo centre for all 25 of the \dorcha~haloes. The different line corresponds to the redshift at which the particles were tagged. \textit{Takeaway:} Most of the tagged particles end up near the halo centre and this is more apparent the higher the redshift of tagging is.}
    \label{fig:solar}
\end{figure}

Our working assumption is that this material traces the phase space of the sites of high redshift star formation. \help{We remind the reader that the correspondence between old stars and metal-poor stars is not one-to-one: metal-poor stars can arise from enriched gas after the first star formation episodes, and stellar age is more reliably inferred from formation time than from metallicity alone.}
In Figure \ref{fig:solar}, we quantify the fraction of material that was in place at $5\leq\,z\leq\,z_i$ within $r=15~\oneh\rm{kpc}$ ($\sim 0.1R_{200}$) of the centre of each \dorcha~halo and plot the corresponding PDFs based on the 25 haloes in our sample. If we focus on the oldest components ($z\geq15$), the PDFs are sharply peaked at 100 per cent, \help{implying that all of these particles are within $r=15~\oneh\rm{kpc}$ for almost all the \dorcha s. In contrast, for $5\leq z\leq10$, the peak is at 80 and 70 per cents --  there is thus a scatter, albeit a small one, amongst the \dorcha  s as to where these progenitors end up at $z=0$. Nevertheless, the trend is for the distributions of material associated with older components to be found predominantly within $r=15~\oneh\rm{kpc}$ for all the \dorcha s. }

This tells us that the oldest stars should be found within the innermost parts of a MWA's halo, within the galaxy disc, most likely contributing to the nuclear bulge. This would be consistent with the kinematics of the $z\geq10$ components in our \dorcha~haloes, which are in line with estimates of the Milky Way's bulge (cf. velocity dispersion of $\sim 100 {~\rm km~s}^{-1}$, \citealt{valentiCentralVelocityDispersion2018}). It is also suggested by the mass density profile (cf. Figure~\ref{fig:radial_profiles}), which drops off as $r^{-4}$, although the amount of mass contained in these components is likely to be negligible. 
\begin{figure}
    \centering
    \includegraphics[width=\columnwidth]{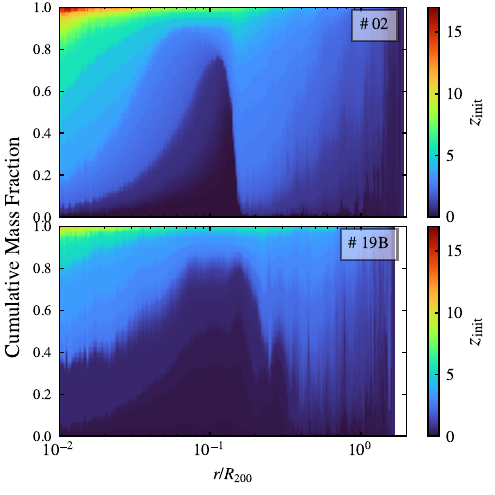}
    \caption{\textbf{Cumulative mass fraction of the progenitor cores:} \help{
    We show the cumulative mass fraction as a function of the normalized radial distance, $r/r_{200}$, for \D{02} and \D{19B} at $z=0$, coloured by their formation redshift ($z_{\rm init}$). At each radius, the colour corresponds to the redshift at which a given fraction (from 0 to 1) of the local cumulative mass formed. \textit{Takeaway:} This reveals how the formation times of particles vary with radius -- for instance, most of the mass within inner regions of the halo, say $r/r_{200}\leq5\times 10^{-2}$, are dominated by the particles comprising the low-$z$ (i.e. $z\leq6$) progenitors.}}
\label{fig:ages_r}
\end{figure}

\help{
\resp{To validate our last assertion}, in Figure \ref{fig:ages_r}, we show the cumulative mass fraction of  the progenitors for \D{02} and \D{19B}\footnote{Note that \D{02} \& \D{19B} are, respectively, the earliest-forming isolated and latest-forming paired \dorcha s. This should help us in maximizing any difference in the trend for our analysis here. We confirm that our discussion here applies to all the \dorcha s.}. For this, we tag and track across all 123 snapshots the 10 per cent most-bound particles of all subhaloes of the progenitor FoF group of our \dorcha s. For every particle, we also keep track of the first redshift ($z_{\rm init}$) at which it forms part of the central core of a progenitor. At $z=0$, we restrict our attention to the main subhalo, and compute the cumulative mass fraction as a function of both radial distance, $r/r_{200}$, as well as $z_{\rm init}$. This allows us to answer: \textit{For a given radius, when did particles that make up a certain fraction of the mass first appear as part of a progenitor?} \resp{It is clear from both subplots of Figure \ref{fig:ages_r} that amongst the progenitor material, only about 20 per cent of the mass in the inner regions ($r/R_{200}\leq5\times10^{-2}$) of the halo are contributed by particles that were in the central cores of progenitors at $z\geq6$. For each of our 25 \dorcha s, we now compute the median $z_{\rm init}$ of tagged particles that lie within $r/R_{200} \le 5 \times 10^{-2}$ at $z=0$. We then take the median of these values. This yields a median-of-medians $z_{\rm init} = 2.25$, with a $1\sigma$ scatter among the individual halo medians of $\Delta z \simeq 0.625$.
}}


\subsection{Our Results in Context}
\label{sec:discussion}

\resp{We have found that the innermost cores of high‑redshift progenitors end up strongly concentrated in the innermost parts of our MWAs by $z$=0, with earlier‑tagged material residing deepest in the potential well. The density profiles of this material fall steeply as $\rho \propto r^{-4}$ and their median $\bar{v}_r \simeq 0$ indicates a cold, dynamically relaxed population long embedded in their host haloes. These trends show a weak dependence on formation time, with only mild variations in velocity anisotropy, $\beta$, and are essentially identical in isolated and paired systems within $R_{200}$. Although individual MWA haloes display shells and caustics, their phase‑space structure is remarkably uniform. Taken together, the remnants of the earliest progenitors form a unified, centrally concentrated, dynamically cold inner component at the present day, largely insensitive to the detailed assembly history of the host. How do these findings compare to previous work?}

\medskip

As we might expect, semi-analytic particle-tagging studies \citep[e.g.][]{cooperGalacticStellarHaloes2010,deasonEatingHabitsMilky2016} make predictions that are consistent with ours - that the oldest stars will be concentrated in the innermost parts of the halo, which can be traced to the importance of a few relatively massive accretion events at early times. Hydrodynamical galaxy formation results from APOSTLE \citep{starkenburgOldestMostMetalpoor2017} show the same trend: the \emph{oldest} stars are preferentially found in the innermost regions of Milky Way–mass galaxies, even though the most metal-poor stars span a wider range of formation times and are not uniquely central. Our finding that this old material follows a steep radial profile, $\rho\propto r^{-4}$, and exhibit $\bar v_r\simeq 0$, complements the APOSTLE prediction \citep{geninaEdgeRelationStellar2023} that the outer halo exhibits phase–space caustics and sharp features that could be traced observationally \citep[see also][]{amoriscoAccretedStellarHalo2017}. The oldest stars, accreted earliest, will follow low velocity dispersion, high–$\beta$ orbits that build a central dynamically cold component, whereas later accreted material is dynamically hotter and at larger radii \citep{cooperGalacticStellarHaloes2010,deasonEatingHabitsMilky2016,amoriscoAccretedStellarHalo2017,fattahiTaleTwoPopulations2020,geninaEdgeRelationStellar2023}.

The APOSTLE simulations find that very metal‐poor and very old stars are a small fraction of the inner population, even if their absolute numbers are astrophysically interesting; the fraction increases strongly with radius and varies with metallicity assumptions \citep{starkenburgOldestMostMetalpoor2017}. The FIRE‐2 simulations show that at fixed $[\mathrm{Fe/H}]$, earlier accretion and/or more massive progenitors have higher $[\alpha/\mathrm{Fe}]$, while later/lower–mass events sit at lower $[\alpha/\mathrm{Fe}]$ \citep{hortaProtogalaxyMilkyWaymass2023}. This is consistent with our finding that the oldest material is dynamically central and kinematically cold/radial: the oldest stars are expected to be preferentially $\alpha$–enhanced, although as other work has shown \citep[e.g.][]{starkenburgOldestMostMetalpoor2017} this may be necessary but not sufficient. However, the work of \citet{reyMEGATRONHowFirst2025a} indicates the identification by metallicity is not straightforward - the VINTERGATAN‑GM simulations show that a radial inner halo at $z=0$ can arise from different mixtures of ex‑situ, accreted, stars and in‑situ star formation, yielding overlapping chemo‑kinematics \citet{reyMEGATRONHowFirst2025a}.

Work based on the FIRE2 simulations indicate that the main progenitor of an Milky Way or M31‑mass galaxy typically emerges by $z\,\sim\,3$–$4$, and that the $z\,=\,0$ bulge (inner $\sim$2\,kpc) formed primarily within a single dominant progenitor by $z\,\lesssim\,5$. In Local‑Group–like pairs, this emergence happens significantly earlier (by $\Delta z\,\sim\,2$), with $\sim4\times$ higher stellar mass at $z\,\gtrsim\,4$. This indicates that environment is important at at early times while remaining consistent with our result that paired and isolated hosts look similar inside $R_{200}$ at $z=0$ \citep{santistevanFormationTimesBuilding2020}.

\section{conclusions}
\label{sec:conclusions}
Galactic archaeology -- searching for fossil signs of early galaxy formation and evolution in the Local Universe --  has the potential to help constrain the first epochs of star formation. By identifying the descendants of the first stars in and around the Galaxy, we can get an understanding of the integrated  interaction history of these objects with the Milky Way as well as how and where they assembled in the early Universe. A number of surveys are ongoing and/or are planned to exploit the vast data associated with individual stars. It is important then that appropriate simulations are on-hand to help interpret them. Moreover, simulations can help design better survey parameters -- for instance,  random blind surveys have a low probability of finding these low-metallicity stars \citep{nessARGOSIIIStellar2013}, particularly in the crowded and dust-obscured inner parts of the MW \citep[][]{howesEMBLASurveyMetalpoor2016, luceyCOMBSSurveyChemical2019, arentsenPristineInnerGalaxy2020}. It is therefore vital to identify any spatial, kinematical, and/or dynamical signatures that these old components may exhibit. 

Motivated by these considerations, we have introduced the dark-matter-only \dorcha~cosmological simulation suite of Milky Way analogues (MWAs). The \dorcha~haloes have been chosen to have masses similar to that of the DM halo of the Milky Way. The suite currently consists of 20 simulations -- 15 \dorcha s are in isolated environments (\D{01} through \D{15}) and the rest are in pairs (\D{16} through \D{20}) giving a total of 25 MWA haloes. Though we are agnostic to the particular merger histories of the MWAs while selecting them for resimulation from the parent simulation, (\D{00})  our isolation criteria results in them having a relatively smooth merger history. Nevertheless, the \dorcha s exhibit a wide range of formation redshifts, from $z_{\rm form} = 0.28$ to $2.22$.

We identify, tag, and track the most-bound 10 per cent particles -- which trace the deepest parts of the gravitational potential well -- for all  high-$z$ progenitor subhaloes of our \dorcha~haloes down from redshifts $(z\in[25, 20, 15, 10, 8, 5])$. We then study and explore the spatial and kinematical distributions of these tagged set of particles in the present day $z=0$ \dorcha s. This include radial profiles (such as the density, mean radial velocity, velocity dispersion, velocity anisotropy, and rotational velocity) as well as their phase space structure. We further sub-categorize the haloes into early- and late-\dorcha s based on their formation redshift $z_{\rm form}$. This enables us to explore if there is any potential impact of recent mergers and/or accretion on our conclusions -- in that a halo that formed most of its mass much earlier than average would relatively have had less disruptive mergers and/or accretion compared to one that formed much later. For $z\leq5$, we track the orbits and compute the minimum pericentric distance for all the ``old'', i.e. $z\geq10$, tagged particles . This enables us to constrain the amount of interactions that these progenitor sites would have had with the central regions of the main subhalo. Further, we study the spatial location of these progenitor sites near the solar neighbourhood. For this we estimate the fraction of these particles that are within $15~\oneh$ kpc of the centre of the halo. \help{We also study the cumulative mass fraction of the \dorcha~haloes at $z=0$ that is composed of the progenitor particles as a function of both radial distance as well as formation redshift}.

\smallskip

\noindent The main conclusions from our analyses are 
\begin{itemize}
    \item From the radial profiles, we show that the particles that constituted the central regions of the high-$z$ progenitors of our \dorcha~haloes are centrally concentrated \resp{(in the main subhalo, at present day)}, mostly within the virial radius $R_{\rm 200}$. This effect is more pronounced the higher the redshift of tagging is. 
    \item The radial density profiles of these progenitors fall off as $\sim r^{-4}$ \citep{hernquistAnalyticalModelSpherical1990}, signifying that these components are centrally concentrated. The median behaviour for these components is $\bar{v}_{\rm r}\simeq0$ km s$^{-1}$ which we interpret as them having been the part of the larger halo for at least a few dynamical times.
    \item We do not find any significant change when we repeat our analyses for the early- and late-\dorcha s --  except for a mild variation for the velocity anisotropy profile $\beta(r)$.This is surprising as we had used the formation redshift, $z_{\rm form}$, as a proxy for the amount of recent accretion/mergers that a halo would have experienced. There are two likely reasons for this: our halo-selection criteria has resulted in relatively quiet growth histories for the \dorcha. Second, even at $z_{50}$ or even $z_{10}$, a typical MWA is $\sim 10^{10}\oneh\Msun$. The main subhaloes of such systems are robust to external perturbations.
    \item We see this notably when we compare the results for isolated and paired \dorcha s. Both of these categories exhibit similar trends. In fact, they are identical within the virial radius $R_{200}$ \citep[see also][]{garrison-kimmelELVISExploringLocal2014}. This is good news if one is mostly concerned with simulations of MWAs; an expensive simulation of two paired haloes can be avoided in this case.
    \item The \dorcha s exhibit unique features in their phase space structures; but the overall trends are very similar. There are well-defined caustics, and distinct shells that reflect materials from different accretion and/or merger events.
    \item Interpreting the radial profiles alongside the phase space structures, we establish that, regardless of the details of the assembly history, the remnants of the high-$z$ progenitors are centrally concentrated in the innermost parts of the potential at z= 0, and dynamically relaxed and cold.
\end{itemize}

We equate the central cores -- traced by the 10 per cent most-bound particles -- of these gravitationally bound DM structures as sites where gas can cool down efficiently to trigger start formation. Our results suggest that all such star formation sites and the stellar material associated with them would have been accreted by the main subhalo of the MWAs. By $z=0$ they would have sunk to the centre of the potential. Likely locations for metal-free PopulationIII stars and their remnants as well as low-metallicity stars that would have formed from their supernova ejecta are thus the central cores of the MWAs. We stress, however, that the correspondence between old stars and metal-poor stars is not one-to-one: metal-poor stars can arise from enriched gas after the first star formation episodes.

As we noted in \S~\ref{sec:intro}, the \dorcha~suite provides the foundation for further study with the \textsc{Solas}~suite of hydrodynamical galaxy formation simulations. We expect that the inclusion of baryons, and the impact of radiative cooling and explosive forms of feedback, will modify the gravitational potential, as well as influencing which progenitors might host star formation and the degree to which they might remain as coherent substructure in phase space. However, we argue that the \dorcha~suite is sufficient to provide robust insights into key structural and kinematic properties of the old stellar population of the Milky Way and its relationship to galaxy formation in the early Universe. We will revisit this in forthcoming papers.


\section*{Acknowledgements}
We thank the anonymous referee for their constructive comments and feedback that helped improve the quality of this work. This research was supported by the Australian Research Council Centre of Excellence for All Sky Astrophysics in 3 Dimensions (ASTRO 3D), through project \#CE170100013. This work was also undertaken with the assistance of resources from the National Computational Infrastructure (NCI Australia), an NCRIS-enabled capability supported by the Australian Government. SB acknowledges the Dr Albert Shimmins Fund for a Postgraduate Writing-Up Award, and  support from grant PID2022-138855NB-C32 funded by MICIU/AEI/10.13039/501100011033 and ERDF/EU, and  project PPIT2024-31833, cofunded by EU--Ministerio de Hacienda y Función Pública--Fondos Europeos--Junta de Andalucía--Consejería de Universidad, Investigación e Innovación.

\smallskip

\textit{Software citations}: We thank the \textsc{MUSIC}, \textsc{AREPO} and \textsc{GADGET-4} development teams for making these codes public. This research also relies heavily, and with great thanks, on the \textsc{python} \citep{PYTHON} open source community, in particular, \textsc{numpy} \citep{NUMPY}, \textsc{matplotlib} \citep{MATPLOTLIB}, \textsc{scipy} \citep{SCIPY}, \textsc{h5py}, \textsc{jupyter} \citep{Jupyter}, and \textsc{pandas} \citep{Pandas}. The visualizations in Figures \ref{fig:parent1} \& \ref{fig:parent2} were generated with \textsc{YT} \citep{turkYtMULTICODEANALYSIS2011}.

\section*{Data Availability}
\dorcha~data products, in the form of halo catalogues and merger trees, are made publicly available on Zenodo \citep[see][]{baluDORCHASimulationsHalo2026}.
Further details can be accessed at \href{https://solas-sims.github.io/data_products/}{https://solas-sims.github.io/data\_products}. 

\bibliography{references}

\appendix
\section{Phase-space structure of the paired \dorcha~haloes}
\label{sec:appendix1}
\begin{figure*}
    \includegraphics[width=1.01\textwidth]{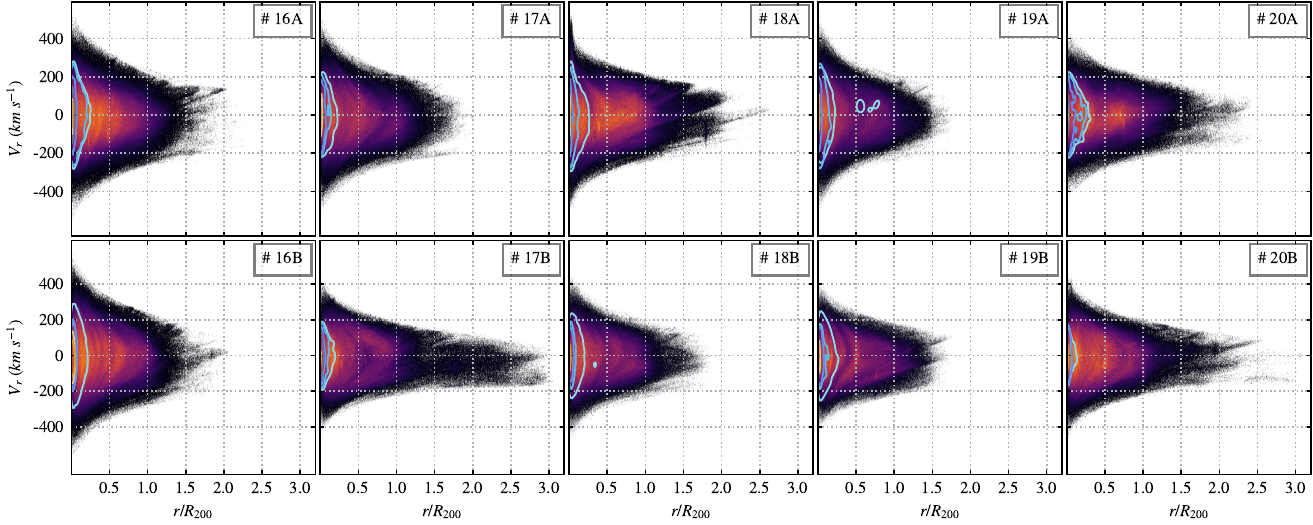}
    \caption{\textit{Phase plot for the paired \dorcha s}: Same as the Figure \ref{fig:phase_plot_iso} but for the paired \dorcha~simulations (\D{15-20}). The more massive of the pair (at $z=0$) is labelled \texttt{A} is shown in the top-row. \textit{Takeaway:} We see a very similar trend for the paired \dorcha s relative to the isolated ones. Here as well, it is clear that most of the tagged particles are concentrated towards the halo centre.}
    \label{fig:phase_plot_pairs}
\end{figure*}

\begin{figure*}
    \includegraphics[width=1.01\textwidth]{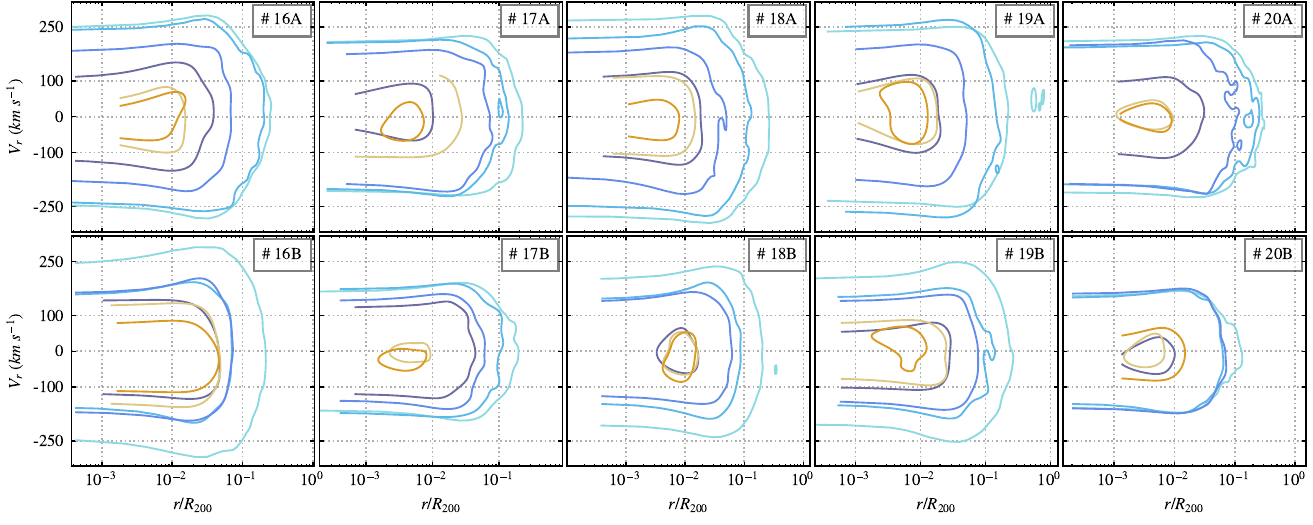}
    \caption{\textit{Phase plot for the isolated \dorcha s}: Shown are the phase-space distribution of entire halo, i.e. the radial velocity $(v_{\rm r})$ against the radial distance $(r)$). The contours represent the 90 per cent contour lines that encompass the phase-space distribution of the tagged particles. \textit{Takeaway:} Figure shows the variety of phase-space distributions of the \dorcha~haloes. It is clear that most of the tagged particles end up within $\sim100\oneh$kpc of the halo centre which is much lower than the typical $R_{200}$.}
    \label{fig:phase_plot_pairs_contours}
\end{figure*}
In Figures \ref{fig:phase_plot_pairs} and \ref{fig:phase_plot_pairs_contours} we show the phase-space structure of the paired \dorcha s (\D{16} to \D{20}). Figure \ref{fig:phase_plot_pairs} is similar to figure \ref{fig:phase_plot_iso} where we show the full phase-space information of all the particles that comprise the corresponding \dorcha~halo. Similar to the isolated \dorcha s (see \ref{ssec:phase_space}) almost all of the particles that we tag at high-$z$ end up concentrated towards the centre of the halo. This is brought out clearly in figure \ref{fig:phase_plot_pairs_contours} where, just like in figure \ref{fig:phase_plot_iso_contours}, we focus on the central portions of the halo and show the 90 percentile contour lines of the distributions of the tagged particles.





\bsp
\label{lastpage}
\end{document}